\documentclass[preprint,12pt,3p]{elsarticle}
\usepackage{graphicx}
\usepackage{amssymb}
\usepackage{caption}
\usepackage{subcaption}
\usepackage{url}
\usepackage{amsmath}
\usepackage{float}

\graphicspath{{./images/}}

\usepackage{tikz}
\tikzset{
  treenode/.style = {shape=rectangle, rounded corners,
                     draw, align=center,
                     top color=white, bottom color=blue!20},
  root/.style     = {treenode, font=\Large, bottom color=red!30},
  env/.style      = {treenode, font=\ttfamily\normalsize},
  dummy/.style    = {circle,draw}
}

\begin{document}

\begin{frontmatter}

\title{Fluid Dynamics in Heart Development: \\ Effects of Hematocrit and Trabeculation}\tnotetext[label0]{This is only an example}

\author[label1]{Nicholas A. Battista\corref{cor1}\fnref{label2}}
\address[label1]{Department of Mathematics, CB 3250, University of North Carolina, Chapel Hill, NC, 27599}
\address[label2]{Department of Biology, 3280, University of North Carolina, Chapel Hill, NC, 27599}

\cortext[cor1]{I am corresponding author}

\ead{nick.battista@unc.edu}
\ead[url]{battista.web.unc.edu}

\author[label1,label3]{Andrea N. Lane}
\address[label3]{Department of Biostatistics, UNC Gillings School of Global Public Health, Chapel Hill, NC, 27599}
\ead{anlane@live.unc.edu}

\author[label4,label5]{Jiandong Liu}
\address[label4]{McAllister Heart Institute, UNC School of Medicine, University of North Carolina, Chapel Hill, NC 27599}
\address[label5]{Department of Pathology and Laboratory Medicine, University of North Carolina, Chapel Hill, NC 27599}


\author[label1,label2]{Laura A. Miller}

\ead{lam9@unc.edu}

\begin{abstract}
Recent \emph{in vivo} experiments have illustrated the importance of understanding the hemodynamics of heart morphogenesis. In particular, ventricular trabeculation is governed by a delicate interaction between hemodynamic forces, myocardial activity, and morphogen gradients, all of which are coupled to genetic regulatory networks. The underlying hemodynamics at the stage of development in which the trabeculae form is particularly complex, given the balance between inertial and viscous forces. Small perturbations in the geometry, scale, and steadiness of the flow can lead to changes in the overall flow structures and chemical morphogen gradients, including the local direction of flow, the transport of morphogens, and the formation of vortices. The immersed boundary method was used to solve the fluid-structure interaction problem of fluid flow moving through a two chambered heart of a zebrafish (\emph{Danio rerio}), with a trabeculated ventricle, at $96\ hpf$ (hours post fertilization). Trabeculae heights and hematocrit were varied, and simulations were conducted for two orders of magnitude of Womersley number, extending beyond the biologically relevant range ($0.2$ -- $12.0$). Both intracardial and intertrabecular vortices formed in the ventricle for biologically relevant parameter values. The bifurcation from smooth streaming flow to vortical flow depends upon the trabeculae geometry, hematocrit, and $Wo$. This work shows the importance of hematocrit and geometry in determining the bulk flow patterns in the heart at this stage of development. 

\end{abstract}


\end{frontmatter}

%
%

\section{Introduction}
\label{Introduction}

Fluid dynamics is important to organogenesis in many systems. The advection and diffusion of morphogens as well as the hemodynamic forces generated are known to regulate morphogenesis \cite{Patterson:2005}. Forces such as shear stress and pressure may be key components that activate developmental regulatory networks \cite{Tarbell:2005}. These mechanical forces act on the cardiac cells, where the mechanical stimuli is then transmitted to the interior of the cell via intracellular signalling pathways, i.e., mechanotransduction. In terms of mixing, the magnitude, direction, and pulsatile behavior of flow near the endothelial layer may influence receptor-ligand bond formation~\cite{Taylor:1996} and enhance the mixing of chemical morphogens. Advection-driven chemical gradients act as epigenetic signals driving morphogenesis in ciliary-driven flows~\cite{Cartwright:09,Freund:12}, and it is possible that flow-driven gradients near the endothelial surface layer also play a role in cardiogenesis and vasculogenesis.

The notion that flow is essential for proper vertebrate cardiogenesis is not a recent idea. It was first investigated by Chapman in 1918 when chicken hearts were surgically dissected during embryogenesis, and their resulting circulatory systems did not develop properly \cite{Chapman:1918}. Moreover, the absence of erythrocytes at the initiation of the first heart beat and for a period of time later, supports the belief that the early developing heart does not pump for nutrient transport. These results suggest that the function of the embryonic heart is to aid in its own growth as well as that of the circulatory system \cite{Burggren:2004}.

Later experiments show that obstructing flow in the venous inflow tract of developing hearts \emph{in vivo} results in problems in proper chamber and valve morphogenesis \cite{Gruber:2004,Hove:2003,Stekelenburg:2008}. For example, Gruber \textit{et al.} \cite{Gruber:2004} found that irregular blood flow can lead to hypoplastic left heart syndrome (HLHS), where the ventricle is too small or absent during the remainder of cardiogenesis. Hove \textit{et al.} \cite{Hove:2003} observed that when inflow and outflow tracts are obstructed in 37 hpf zebrafish, regular waves of myocardial contractions continue to persist and neither valvulogenesis, cardiac looping, nor chamber ballooning occur. Similarly, de-Vos \textit{et al.} \cite{Stekelenburg:2008} performed a similar experiment in HH-stage 17 chicken embryos whereby the venous inflow tract was obstructed temporarily. They noticed that all hemodynamic parameters decreased initially, i.e., heart rate, peak systolic velocity, time-averaged velocity, peak and mean volumetric flow, and stroke volume. Only the heart rate, time-averaged velocity, and mean volumetric flow recovered near baseline levels. 



Trabeculae are particularly sensitive to changes in intracardiac hemodynamics \cite{Granados:2012}. Trabeculae are bundles of muscle that protrude from the interior walls of the ventricles of the heart. The sensitivity of the trabeulae under varying mechanical loads is important when considering they may serve as important structures in which cellular mechanotransduction occurs. Trabeculation may also help regulate and distribute shear stress over the ventricular endocardium, enhance mixing, and modify chemical morphogen gradients. Furthermore, the presence of trabeculation may contribute to a more uniform transmural stress distribution over the cardiac wall \cite{Malone:2007}. Even subtle trabeculation defects spawning from slight modifications in hemodynamics may magnify over time. As the mechanical force distribution changes due to the absence of normal trabeculae, Neuregulin signalling, along with other genetic signals, are disrupted, leading to further deviations from healthy cardiogenesis. For example, zebrafish embryos that are deficient in the key Neuregulin co-receptor ErbB2 display severe cardiovascular defects including bradycardia, decreased fractional shortening, and impaired cardiac conduction \cite{Liu:2010}. Disrupted shear distributions in the ventricle leads to immature myocardial activation patterns, which perpetuate ventricular conduction and contractile deficiencies, i.e., arrhythmia, abnormal fractional shortening, and possibly ventricular fibrillation \cite{Reckova:2003}. 

The fluid dynamics of heart development, particularly at the stage when the trabeculae form, is complex due to the balance of inertial and viscous forces. The Reynolds number, $Re$ is a dimensionless number that describes the ratio of inertial to viscous forces in the fluid and is given as $Re = (\rho U L)/\mu$. In cardiac applications, $\mu$ is the viscosity of the blood, $\rho$ is the density of the blood, $U$ is the characteristic velocity (often chosen as the average or peak flow rate), and $L$ is the characteristic length (often chosen as the diameter of the chamber or vessel). Another dimensionless parameter that is often used in describing cardiac flows is the Womersley number which is given by $Wo = \frac{\omega L^2}{\nu}$, where $\omega$ is the angular frequency of contraction. Note that the $Wo$ describes the transient inertial force over the viscous force and is a measure of the importance of unsteadiness in the fluid. During critical developmental stages such as cardiac looping and the formation of the trabeculae, $Re \approx 1$ and $Wo \approx 1$. In this regime, a number of fluid dynamic transitions can occur, such as the onset of vortical flow and changes in flow direction, that depend upon the morphology, size of the chambers, and effective viscosity of the blood. The flow is also unsteady, and the elastic walls of the heart undergo large deformations. 

Since analytical solutions are not readily available for complex geometries at intermediate $Re$, recent work has used computational fluid dynamics to resolve the flow in the embryonic heart. For example, DeGroff et al. \cite{DeGroff:2003} reconstructed the three-dimensional surface of human heart embryo using a sequence of two-dimensional cross-sectional images at stages 10 and 11. The cardiac wall was fixed, and steady and pulsatile flows were driven through the chambers. They found streaming flows (particles released on one side of the lumen did not cross over or mix with particles released from the opposite side) without coherent vortex structures. Liu et al. \cite{Liu:2007} simulated flow through a three-dimensional model of a chick embryonic heart during stage HH21 (after about 3.5 days of incubation) at a maximum $Re$ of about 6.9. They found that vortices formed during the ejection phase near the inner curvature of the outflow tract. More recently, Lee \textit{et al.} \cite{Lee:2013} performed 2D simulations of the developing zebrafish heart with moving cardiac walls. They found unsteady vortices develop during atrial relaxation at 20-30 hpf and in both the atrium and ventricle at 110-120 hpf. Goenezen \textit{et al.} \cite{Goenezen:2015} used subject-specific computational fluid dynamics (CFD) to model flow through a model of the chick embryonic heart outflow tract.

The numerical work described above, in addition to direct \textit{in vivo} measurements of blood flow in the embryonic heart \cite{Hove:2003,Vennemann:2006}, further supports that the presence of vortices is sensitive to changes in $Re$, morphology, and unsteadiness of the flow. Santhanakrishnan \textit{et al.} \cite{Santhanakrishnan:2009} used a combination of CFD and flow visualization in dynamically scaled physical models to describe the fluid dynamic transitions that occur as the chambers balloon, the endocardial cushions grow, and the overall scale of the heart increases. They found that the formation of intracardial vortices depended upon the height of the endocardial cushions, the depth of the chambers, and the $Re$. Their paper only considers steady flows in an idealized two-dimensional chamber geometry with smooth, stationary walls.

In this paper, we quantify the kinematics of the two-chambered zebrafish heart at $96\ hpf$ and use that data to construct both the geometry and prescribed pumping motion of a two-chambered heart computational model.  We use the immersed boundary method to solve the fluid-structure interaction problem of flow through a two-chambered pumping heart. The goal of this paper is to discern bifurcations in the intracardial and intertrabecular flow structures due to scale ($Wo$), trabeculae height, and hematocrit. We find a variety of interesting bifurcations in flow structures that occur over a biologically relevant morphospace. The implications of the work are that alterations in bulk flow patterns, and particularly the presence or absence of intracardial and intertrabecular vortices, will augment or reduce mixing in the heart, alter the direction and magnitude of flow near the endothelial surface layer, and potentially change chemical gradients of morphogens which serve as an epigenetic signal.



%
%

\section{Numerical Method}
\label{numerical_section}

\subsection{Model Geometry}
\label{Model_Geometry_sub}

%
%

\begin{figure}
    \begin{subfigure}{0.6\linewidth}
        \centering
        \includegraphics[scale=1.0]{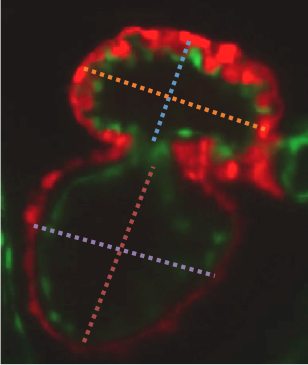}
        \caption{}
        \label{ZF_Vcon}
    \end{subfigure}
    \begin{subfigure}{0.35\linewidth}
        \includegraphics[scale=1.0]{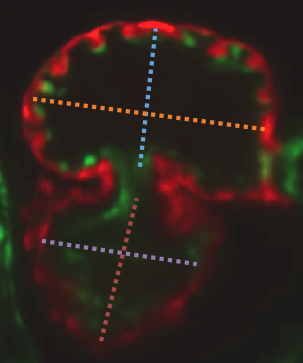}
        \caption{}
        \label{ZF_Vexp}
    \end{subfigure}\\ \\
    \begin{subfigure}{0.6\linewidth}
        \centering
        \includegraphics[scale=0.62]{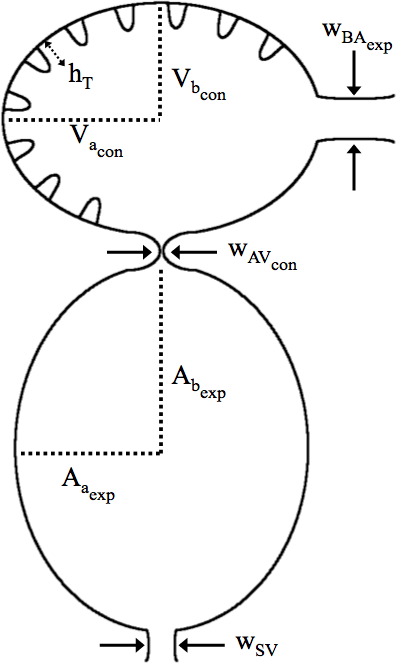}
        \caption{}
        \label{Geo_Vcon}
    \end{subfigure}
    \begin{subfigure}{0.35\linewidth}
        \includegraphics[scale=0.62]{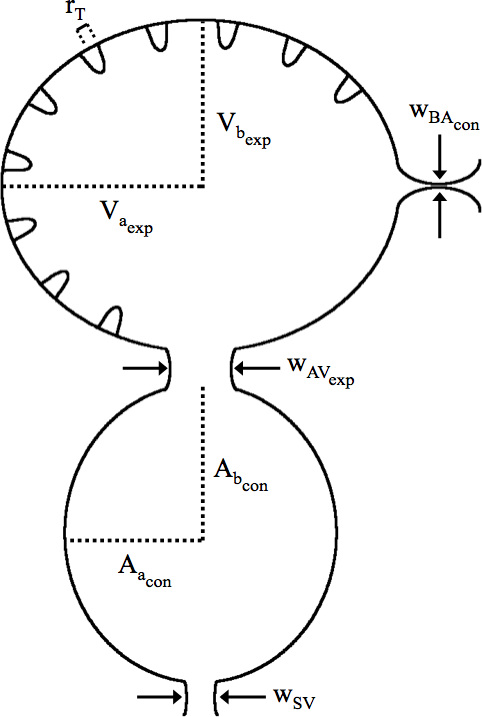}
        \caption{}
        \label{Geo_Vexp}
    \end{subfigure}\\ \\
    \begin{subfigure}{\linewidth}
        \centering
        \includegraphics[scale=1.0]{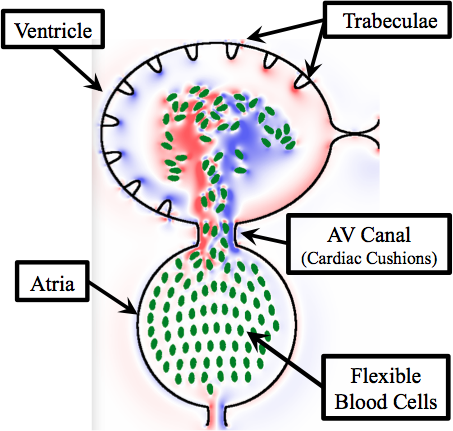}
        \caption{}
        \label{Geometry}
    \end{subfigure}
\caption{\ref{ZF_Vcon} and \ref{ZF_Vexp} are snapshots of an embryonic zebrafish's ventricle at 96 hpf right using spinning disk confocal microscopy. The snapshots were taken right before its diastolic and systolic phase, respectively. The protrusions into the ventriclular chamber are trabeculae. Dashed lines show the minor and major axes. Images are from Tg(cmlc2:dsRed)s879; Tg(flk1:mcherry)s843 embryos expressing fluorescent proteins that label the myocardium and endocardium, respectively \cite{Liu:2010}.  \ref{Geo_Vcon} and \ref{Geo_Vexp} illustrate the computational geometry right before diastole and systole, respectively. The computational geometry, as shown in \ref{Geometry}, includes the two chambers, the atria (bottom chamber) and ventricle (top chamber), the atrioventricular canal connecting the chambers, and the bulbus arteriosus and sinus venosus, which all have endocardial cushions, which can occlude cardiac flow, as well as flexible blood cells.}
\label{Model_Geometry}
\end{figure}

A simplified two dimensional geometry of a 96 hpf zebrafish's two-chambered heart, containing trabeculae, was constructed using Figure \ref{ZF_Vcon} and \ref{ZF_Vexp}. The ventricle and atria were idealized as an ellipse, with semi-major axis $V_a$, and $A_a$, and semi-minor axis $V_b$ and $A_b$, respectively. The atrioventricular canal (AV canal) connects the atria and ventricle and is modeled as endocardial cushions, which move to occlude or promote flow through the heart chambers. The sinus venosus (SV) and bulbus arteriosus (BA) are modeled similarly. The width of the AV canal, SV, and BA are given by $w_{AV}, w_{SV},$ and $w_{BA}$, respectively. The above parameters are labeled systole and diastole separately, e.g., the ventricular subscripts are given an $exp$ label right before systole and are labeled $con$ before diastole, while the atrial labels are opposite.

\begin{figure}[H]
    \begin{subfigure}{0.6\linewidth}
        \centering
        \includegraphics[scale=1.4]{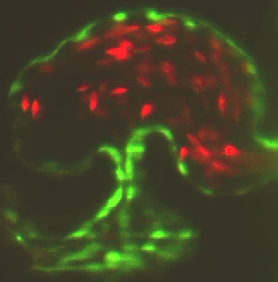}
        \caption{}
        \label{ZF_BCs_exp}
    \end{subfigure}
    \begin{subfigure}{0.35\linewidth}
        \includegraphics[scale=1.4]{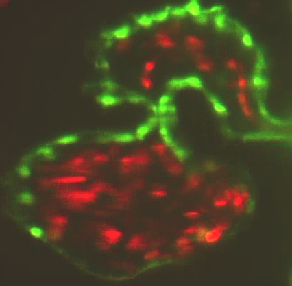}
        \caption{}
        \label{ZF_BCs_con}
    \end{subfigure}\\ \\
\caption{\ref{ZF_BCs_exp} and \ref{ZF_BCs_con} are snapshots of an embryonic zebrafish's ventricle at 96 hpf right using spinning disk confocal microscopy. The snapshots were taken right before systole and diastole, respectively. The protrusions into the ventriclular chamber are trabeculae and blood cells are flouresing red \cite{Liu:2010}.}
\label{ZF_BCs_fig}
\end{figure}

Elliptical blood cells of uniform semi-major and semi-minor axis lengths, $C_a$ and $C_b$, respectively., were included The volume fraction, or hematocrit, was varied between $[0\%, 25\%]$. Hemaocrit increases linearly throughout development \cite{Roubaie:2011} from $0\%$ to roughly $32\%$ \cite{Eames:2010}. The desired volume fraction of blood cells was calculated within the atria, and the blood cells were spaced evenly apart within it. Moreover, as the ejection fraction is $60\%$ \cite{Forouhar:2006}, $60\%$ of the number of blood cells in the atrium were spaced evenly within the ventricle. \emph{in vivo} images from \cite{Liu:2010} are shown in Figure \ref{ZF_BCs_fig}.  Note that this placement of blood cells occurred immediately before diastole. 

The trabeculae geometry was modeled using the following perturbed Gaussian-like function,
\begin{equation}
\label{TrabGeometry} T(x) = h_T \left(1 - \left(\frac{x}{r_T}\right)^2 \right) e^{-\left(\frac{x}{0.7r_T} \right)^8 },
\end{equation}
where $r_T$ and $h_T$ are the radii and height of each trabecula, respectively. Trabeculae are placed equidistant apart, as estimated from Figures \ref{ZF_Vcon} and {\ref{ZF_Vexp}.  The full geometry can be seen in Figure(\ref{Model_Geometry}). 

The blood cells were approximated as ellipses, using Figure \ref{ZF_BCs_fig} to estimate their length to width ratios, with respect to the size of the ventricle. The blood cells were held nearly rigid, as described in Section \ref{FluidStructureModelSection}. 

The dimensionless geometric model parameters are found in Table(\ref{Geo_Params}), which were scaled from measurements taken from Figures \ref{ZF_Vcon} and \ref{ZF_Vexp}. The radii, $r_T$, and number of the trabeculae were constant in all numerical simulations, while the height of the trabeculae, $h_T$, was varied.



\begin{table}[]
\centering
\begin{tabular}{|l|c|c|}
\hline
Parameter & Symbol & Value \\
\hline
Contracted Ventricle Semi-Major Axis & $V_{a_{con}}$ & 0.80  \\
Contracted Atria Semi-Major Axis     & $A_{a_{con}}$ & 0.68 \\
Contracted Ventricle Semi-Minor Axis & $V_{b_{con}}$ & 0.64 \\
Contracted Atria Semi-Minor Axis     & $A_{b_{con}}$ & 0.76 \\
Expanded Ventricle Semi-Major Axis   & $V_{a_{exp}}$ & 1.00 \\
Expanded Atria Semi-Major Axis       & $A_{a_{exp}}$ & 0.88 \\
Expanded Ventricle Semi-Minor Axis   & $V_{b_{exp}}$ & 0.84 \\
Expanded Atria Semi-Minor Axis       & $A_{b_{exp}}$ & 1.02 \\
Contracted AV-Canal Width            & $w_{AV_{con}}$& 0.02 \\ 
Contracted Bulbus Arteriosus Width   & $w_{BA_{con}}$ & 0.015 \\ 
Open AV-Canal Width                  & $w_{AV_{exp}}$ & 0.34 \\ 
Open Bulbus Arteriosus Width         & $w_{BA_{exp}}$ & 0.29 \\
Sinus Venosus Width                  & $w_{SV}$ & 0.2 \\
Blood Cell Semi-Major Axis           & $C_a$ & 0.050 \\
Blood Cell Semi-Minor Axis           & $C_b$ & 0.025 \\
Trabeculae Radii                     & $r_T$ & 0.06 \\
Trabeculae Height                    & $h_T$ & \{0, 0.09, 0.18, 0.27, 0.36\}\\
\hline
\end{tabular}
\caption{Table of dimensionless geometric parameters used in the numerical model. The non-dimensionalization was done by dividing by $V_{a_{exp}}$. The height of trabeculae, $h_T$, were varied for numerical experiments.}
\label{Geo_Params}
\end{table}

%
%

\subsection{Numerical Method}
\label{FluidStructureModelSection}

The immersed boundary method \cite{Peskin:2002} was used to solve for the flow velocities within the geometric model from Section \ref{Model_Geometry_sub}. The immersed boundary method has been successfully used to study the fluid dynamics of a variety of biological problems in the intermediate Reynolds number range, defined here as $0.01<Re<1000$ (see, for example, \cite{Jung:2001,Hershlag:2011,Bhalla:2013a,Tytell:2010}). The model consists of stiff boundaries that are immersed within an incompressible fluid of dynamic viscosity, $\mu$, and density, $\rho$. The fluid motion is described using the full 2D Navier-Stokes equations given as

\begin{equation}
\label{Navier_Stokes} \rho \left( \frac{\partial {\bf{u}}({\bf{x}},t) }{\partial t} + {\bf{u}}({\bf{x}},t)\cdot \nabla {\bf{u}}({\bf{x}},t) \right) = -\nabla p({\bf{x}},t) + \mu \Delta {\bf{u}}({\bf{x}},t) + {\bf{F}}({\bf{x}},t) \\
\end{equation}
\begin{equation}
\label{Incompressibility} \nabla\cdot {\bf{u}}({\bf{x}},t) = 0,
\end{equation}
where ${\bf{u}}({\bf{x}},t) = (u({\bf{x}},t),v({\bf{x}},t))$ is the fluid velocity, $p(\bf{x},t)$ is the pressure, ${\bf{F}}({\bf{x}},t)$ is the force per unit volume (area in $2D$) applied to the fluid by the immersed boundary, i.e., two-chambered heart. The independent variables are the position, ${{\bf{x}}}= (x,y)$, and time, $t$. Eq.(\ref{Navier_Stokes}) is equivalent to the conservation of momentum for a fluid, while Eq.(\ref{Incompressibility}) is a condition mandating that the fluid is incompressible. \\

The interaction equations between the fluid and the immersed structure are given by
\begin{equation}
\label{IBM_Force} {\bf{F}}({\bf{x}},t) = \int {\bf{f}}(r,t)\delta({\bf{x}}-{\bf{X}}(r,t)) dr
\end{equation}
\begin{equation}
\label{IBM_Velocity} {\bf{U}}({\bf{X}}(r,t),t) = \frac{\partial {\bf{X}}(r,t)}{\partial t} = \int {\bf{u}}({\bf{x}},t) \delta( {\bf{x}} - {\bf{X}}(r,t) ) d{\bf{x}},
\end{equation}
where ${\bf{X}}(r,t)$ gives the Cartesian coordinates at time $t$ of the material point labeled by Lagrangian parameter $r$, ${\bf{f}}(r,t)$ is the force per unit area imposed onto the fluid by elastic deformations in the boundary, as a function of the Lagrangian position, $r$, and time, $t$. Eq.(\ref{IBM_Force}) applies a force from the immersed boundary to the fluid grid through a delta-kernel integral transformation. Eq.(\ref{IBM_Velocity}) sets the velocity of the boundary equal to the local fluid velocity.

The force equations are specific to the application. In a simple case where a preferred motion or position is enforced, boundary points are tethered to target points via springs. The equation describing the force applied to the fluid by the boundary in Lagrangian coordinates is given by ${\bf{f}}(r,t)$ and is explicitly written as,
\begin{equation}
\label{IBM_force_explicit} {\bf{f}}_{trgt}(r,t) = k_{target} \left( {\bf{Y}}(r,t) - {\bf{X}}(r,t) \right),
\end{equation}
where $k_{target}$ is the stiffness coefficient, and $\bf{Y}(r,t)$ is the preferred position Lagrangian position of the target structure. In all simulations the motion of the two-chambered heart (atria, ventricle, AV canal, SV, and BA) was prescribed by applying a force proportional to the distance between location of the actual boundary and the preferred position. The deviation between the actual and preferred positions can be controlled with the variable $k_{target}$. 

The blood cells' deformations and movement was governed by fully coupled fluid-structure interaction and movement was not prescribed. Linear springs were used to model the flexibility of blood cells; however, the spring stiffnesses were large as only to allow negligible deformations. Springs were attached between both adjacent Lagrangian points as well as the Lagrangian point across from them. The forces applied to the fluid from deformations of the blood cells is given by
\begin{equation}
\label{IBM_force_spring} {\bf{f}}_{spr}(r,t) = k_{spring} \left( 1 - \frac{R_L}{||{\bf{X}}_{SL}(r,t)-{\bf{X}}_{M}(r,t)||} \right) \cdot \left( \begin{array}{c} x_{SL} - x_M \\ y_{SL} - y_M \end{array} \right),
\end{equation} \\
where ${\bf{X}}_{M}$ and ${\bf{X}}_{SL}$ are the master and slave node, respectively, $k_{spring}$ is the spring stiffness, and $R_L$ is the resting length of the spring.  

%
%

\subsection{Prescribed Motion of the Two Chambers}
\label{PrescribedMotion}

\begin{figure}
    \begin{subfigure}{\linewidth}
        \centering
        \includegraphics[scale=0.45]{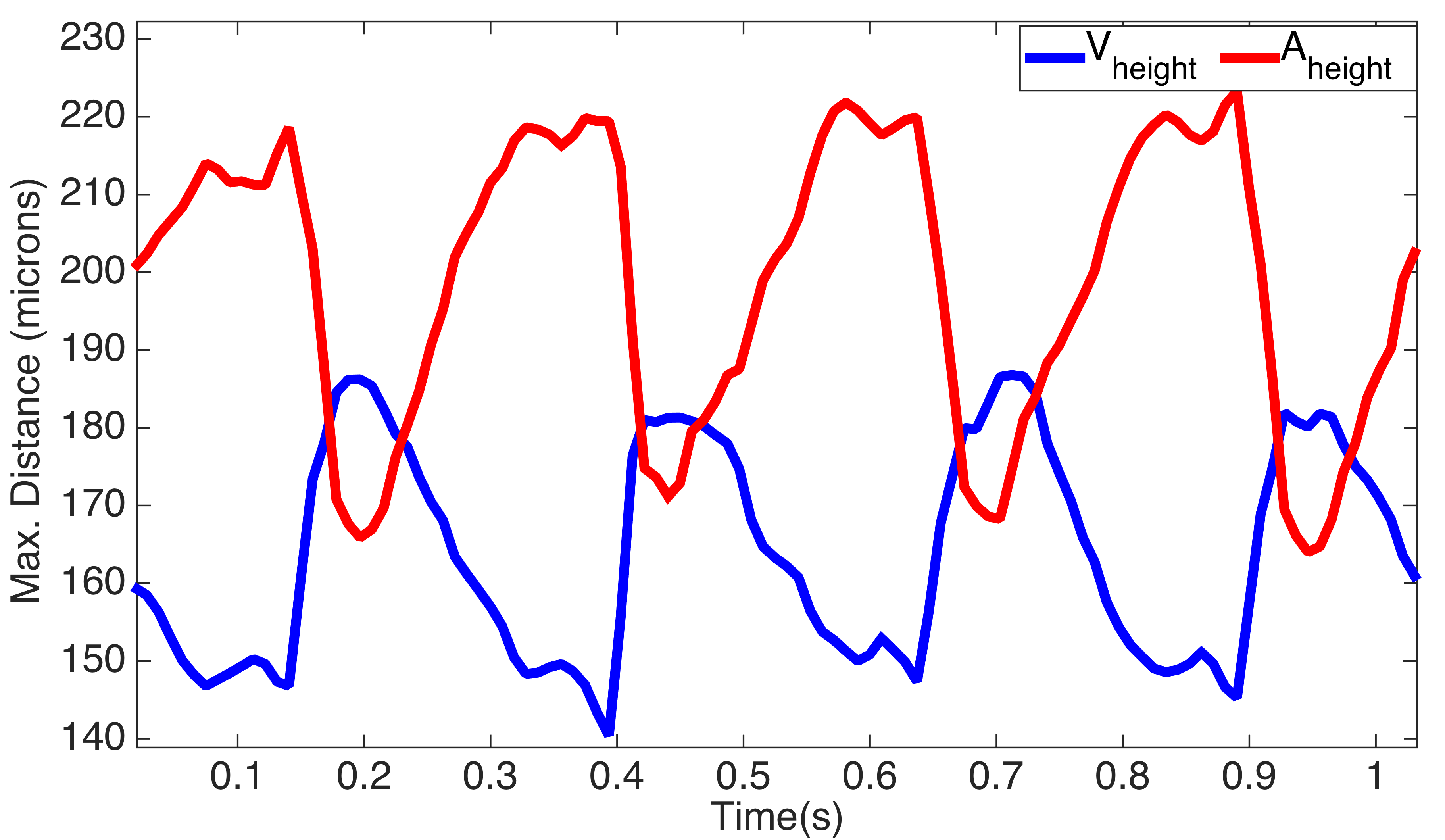}
        \caption{}
        \label{HeightKinematics}
    \end{subfigure} \\ \\
    \begin{subfigure}{\linewidth}
        \centering
        \includegraphics[scale=0.45]{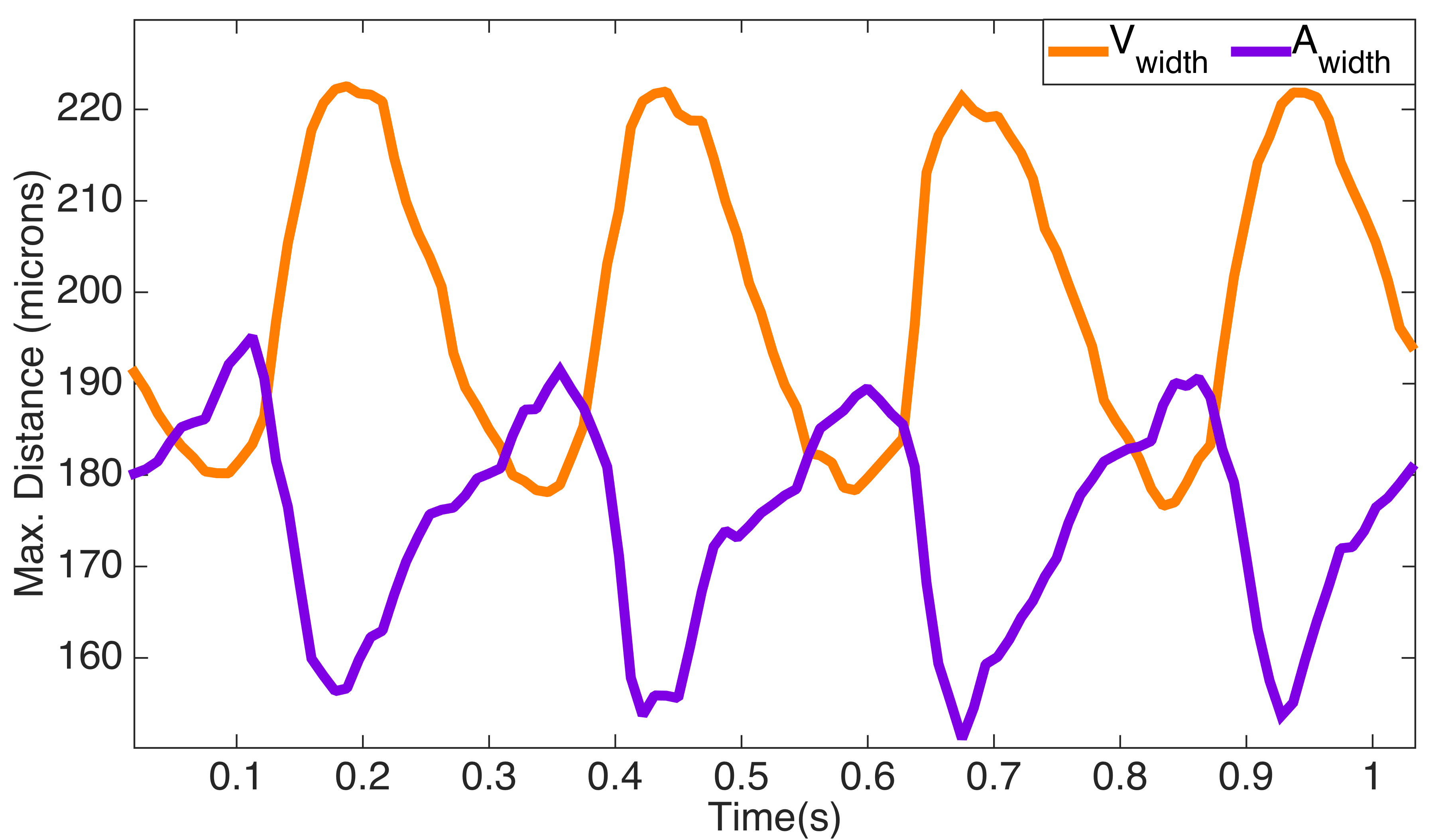}
        \caption{}
        \label{WidthKinematics}
    \end{subfigure}\\ \\
\caption{\ref{HeightKinematics} and \ref{WidthKinematics} illustrate the maximum distance for the height and width (in pixels) respectively, in the atria and ventricle of a $4\ dpf$ embryonic zebrafish heart from \cite{Liu:2010}.}
\label{ZF_BCs}
\end{figure}

\begin{table}[]
\centering
\begin{tabular}{|c|c|c|c|c|c|}
\hline
\multicolumn{2}{|c|}{Ventricular Parameters} & \multicolumn{2}{|c|}{Atrial Parameters} & \multicolumn{2}{|c|}{Trabecular Parameters}\\
\hline
Parameter & Max. Length ($\mu m$) & Parameter & Max. Length ($\mu m$) & Parameter & Length($\mu m$) \\
\hline
$\tilde{V}_{a_{con}}$ & 89.20  & $\tilde{A}_{a_{exp}}$ & 98.11 & $\tilde{r}_T$ & 7.29 \\
$\tilde{V}_{b_{con}}$ & 70.84  & $\tilde{A}_{b_{exp}}$ & 113.11& $\tilde{h}_T$ & 20.97 \\
$\tilde{V}_{a_{exp}}$ & 93.78  & $\tilde{A}_{a_{con}}$ & 76.59 & &\\
$\tilde{V}_{b_{exp}}$ & 111.98 & $\tilde{A}_{b_{con}}$ & 84.10 & &\\
\hline
\end{tabular}
\caption{The morphological parameters in physical units as computed from the kinematic analysis.}
\label{HeightWidthTable}
\end{table}

The motion of the two-chambered heart was modeled after a video taken using spinning disk confocal microscopy from \cite{Liu:2010} of a wildtype zebrafish embryo at $96$hpf. The video's images were acquired with a Nikon Te-2000u microscope (Nikon) at a rate of 250 frames per second using a high-speed CMOS camera (MiCam Ultima, SciMedia) \cite{Liu:2010}. Using the MATLAB software package DLTdv \cite{Hedrick:2008}, the systolic and diastolic periods were determined by measuring maximum width and height of both the atrial and ventricular chambers. These results are shown in Figure \ref{HeightKinematics} and \ref{WidthKinematics}. The maximum width of the AV canal was also measured in pixels right before diastole, and found to be $25$ pixels. Assuming the width of the AV canal is $42 \mu m$ \cite{Forouhar:2006}, each single pixel corresponds to $1.68 \mu m.$ The converted height and widths in $\mu m$ are found in Table \ref{HeightWidthTable}. The average heights and radii of the trabeculae were found to be $20.96\mu m$ and $7.29\mu m$ respectively.

\begin{table}[]
\centering
\begin{tabular}{|c|c|c||c|c|c|}
\hline
\multicolumn{3}{|c||}{Ventricular Phases} & \multicolumn{3}{|c|}{Atrial Phases} \\
\hline
Phase & $\% of Period$ &  Time ($s$) & Phase & $\% of Period$ & Time ($s$) \\
\hline
Rest after Contraction & 20.7 & 0.05 & Rest after Expansion   & 20.7 & 0.05 \\
Expansion          & 24.0 & 0.06 & Contraction        & 24.0 & 0.06 \\
Rest after Expansion   & 7.9  & 0.02 & Rest after Contraction & 2.2  & 0.01 \\
Contraction        & 47.4 & 0.12 & Expansion          & 53.1 & 0.13 \\
\hline
\end{tabular}
\caption{Average percentage and duration of each phase during the heart cycle obtained from kinematic analysis.}
\label{TimesTable}
\end{table}

One entire heart cycle was found to take place in approximately $27$ frames. Assuming the heart beat frequency is $3.95\ beats/s$ \cite{Malone:2007}, each pumping cycle lasts $\sim0.25s$. Each heart chamber undergoes four phases during each cycle: a rest period at the end of contraction, a period of expansion, a rest period at the end of expansion, and a period of contraction. The average percentage and duration of each phase are given in Table \ref{TimesTable}.


\begin{figure}
    \centering
    \includegraphics[scale=0.85]{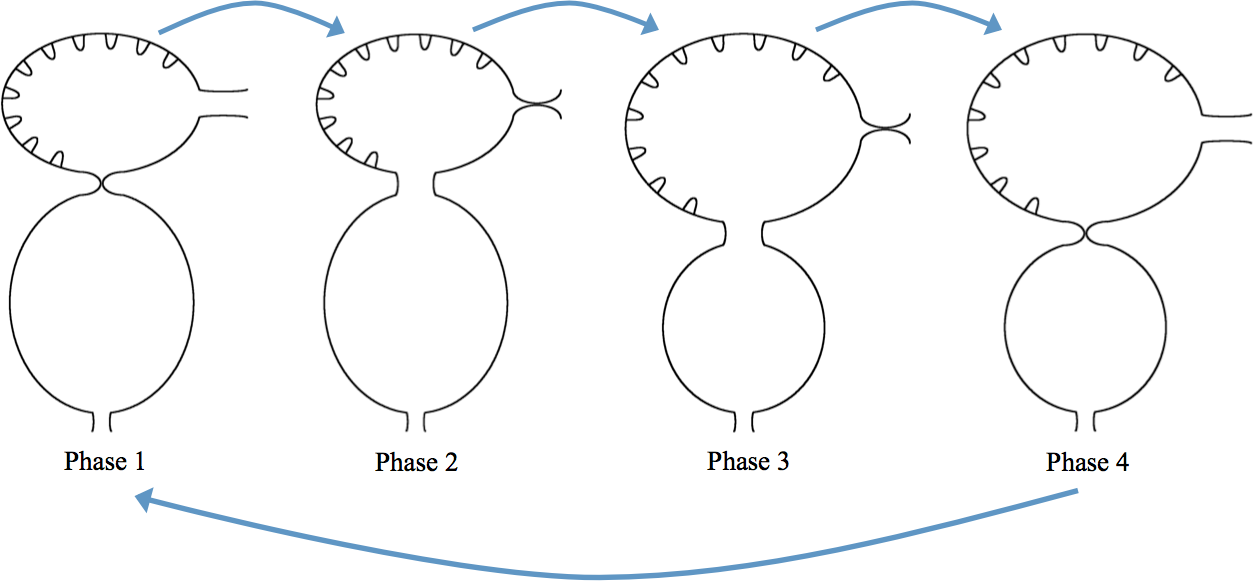}
    \caption{We describe four phases of each heart cycle. Note that the position at the beginning of each phase is shown. Phase 1: the ventricle rests after contraction and the atrium rests after expansion. The AV canal goes from fully occluded to $10\%$ occlusion. Phase 2: The diastolic phase when the ventricle expands while the atria contracts. Phase 3: the ventricle rests after expansion and the atrium rests after contraction. The AV canal becomes fully occluded state. Phase 4: The systolic phase, when the ventricle contracts and the atria expands.}
    \label{Phases}
\end{figure} 

The prescribed motion of the two-chamber hearts was performed by interpolating between different phases of the heart cycle. This is illustrated in Figure \ref{Phases} which show the beginning of each phase. Phase 1: the ventricle rests after contraction and the atrium rests after expansion. The AV canal goes from fully occluded to $10\%$ occlusion. Phase 2: The diastolic phase when the ventricle expands while the atria contracts. Phase 3: the ventricle rests after expansion and the atrium rests after contraction. The AV canal becomes fully occluded state. Phase 4: The systolic phase, when the ventricle contracts and the atria expands. Note we model the time in each phase after the ventricle motion, only. 

The actual motion of the heart is driven by changing the preferred position of the target points. Each phase transition used the following interpolation function, 

$${\bf{X}}_{target} = {\bf{X}}_{current} + g_j(t)\left[ {\bf{X}}_{next} - {\bf{X}}_{current} \right],$$

\noindent where 

\begin{equation}
\label{interp_function} g_j(t) = \left\{ \begin{array}{ll} c_1 \left( \frac{t}{T_{P_j}} \right)^2 & t<t_1 \\
c_3\left(\frac{t}{T_{P_j}}\right)^3 + c_4\left( \frac{t}{T_{P_j}}\right)^2 + c_5\left(\frac{t}{T_{P_j}}\right) + c_6 & t_1\leq t \leq t_2 \\ 
-c_2 \left( \frac{t}{T_{P_j}}-1 \right)^2 + 1 & t>t_2 \end{array}\right.,\\ \\
\end{equation}

\noindent and $T_{P_j}$ is the total time for Phase $j$. Eq. (\ref{interp_function}) was chosen to enforce continuous accelerations between phases. The coefficients $\{c_k\}_{k=1}^{6}$ are given in Table \ref{PolyCoeffs} and the durations of each phase are reported in Table \ref{InterpTime}.

\begin{table}[]
\centering
\begin{tabular}{|c|c|}
\hline
Parameter & Value \\
\hline
$c_1$ & 2.739726027397260 \\
$c_2$ & 2.739726027397260 \\
$c_3$ & -2.029426686960933 \\
$c_4$ & 3.044140030441400 \\
$c_5$ & -0.015220700152207 \\
$c_6$ & 0.000253678335870 \\
\hline
\end{tabular}
\caption{Table of polynomial coefficients for the interpolating function, $g_j(t)$.}
\label{PolyCoeffs}
\end{table}

\begin{table}[]
\centering
\begin{tabular}{|c|c||c|c|}
\hline
\multicolumn{2}{|c||}{Phase 1} & \multicolumn{2}{|c|}{Phase 2}  \\
\hline
Parameter &  Time  & Parameter &  Time  \\
\hline
$T_P$ & $0.207\times$ Period & $T_P$ & $0.240\times$ Period  \\ 
$t_1$ & $0.05\times T_{P_1}$ & $t_1$ & $0.07\times T_{P_2}$  \\
$t_2$ & $0.95\times T_{P_1}$ & $t_2$ & $0.93\times T_{P_2}$  \\
\hline 
\hline
\multicolumn{2}{|c||}{Phase 3} & \multicolumn{2}{|c|}{Phase 4}\\
\hline
Parameter &  Time  & Parameter &  Time \\
$T_P$ & $0.079\times$ Period & $T_P$ & $0.474\times$ Period \\
$t_1$ & $0.05\times T_{P_3}$ & $t_1$ & $0.04\times T_{P_4}$ \\
$t_2$ & $0.95\times T_{P_3}$ & $t_2$ & $0.96\times T_{P_4}$ \\
\hline
\end{tabular}
\caption{Table of temporal parameters used in the interpolating function, $g_j(t)$.}
\label{InterpTime}
\end{table}

To determine the $Wo$ within the heart, we take characteristic values for zebrafish embryonic hearts between $4$ and $4.5$ dpf and match our dimensionless model parameters accordingly. The characteristic frequency, $f_{zf}$ was measured \textit{in vivo}, and the characteristic length, $L_{zf}$, was taken as the height of the ventricular right before systole. The $Wo$ was then calculated as 
\begin{equation}
Wo = L_{zf}\ \sqrt{ \frac{2\pi\cdot f_{zf}\cdot \rho_{zf} }{ \mu_{zf} }  }  = 0.77,
\end{equation}
\noindent where $f_{zf} = 3.95\ s^{-1}$ \cite{Malone:2007}, $\rho_{zf} = 1025$ $kg/m^3$ \cite{Santhanakrishnan:2011}, $\mu_{zf} = 0.0015$ $kg / (m\cdot s)$ \cite{Mohammed:2011}, and $L_{zf} = 0.188\ mm$ from DLTdv analysis. 

The characteristic velocity, $V_{zf}$, was taken as the average of the minimum and maximum velocity measured \textit{in vivo}. The dimensionless frequency may then be calculated as
\begin{equation}
\tilde{f}=\frac{L_{zf}}{V_{zf}}\cdot f_{zf} = 0.1,
\end{equation}
where $V_{zf} = 0.75$ $cm/s$ \cite{Hove:2003}.

For the mathematical model, the parameters were chosen to keep the dimensionless frequency fixed at $\tilde{f}=1.0$. The $Wo$ was varied by changing the kinematic viscosity, $\nu = \mu / \rho$. The computational parameters are reported in Table(\ref{InterpTime}). For the simulations, the $Wo_{sim}$ is calculated using a characteristic length of $V_{b_{exp}}$ and characteristic velocity is set to the maximum velocity in the AV canal during diastole. Since the pumping motion is prescribed, the maximum velocity in the AV canal remains close to constant, regardless of $Wo$. The simulations were performed for $Wo_{sim} =  \{0.2, 0.5,1.0,2.0,4.0,8.0,12.0\}$. The stiffnesses of the target points were chosen the minimize the deviations from the preferred position and varied with $Wo_{sim}$ and the same stiffness was used in all simulations. 

We used an adaptive and parallelized version of the immersed boundary method, IBAMR \citep{BGriffithIBAMR,Griffith:2007}. IBAMR is a C++ framework that provides discretization and solver infrastructure for partial differential equations on block-structured locally refined Eulerian grids \citep{MJBerger84,MJBerger89} and on Lagrangian (structural) meshes. IBAMR also includes infrastructure for coupling Eulerian and Lagrangian representations.  

The Eulerian grid on which the Navier-Stokes equations were solved was locally refined near the immersed boundaries and regions of vorticity with a threshold of $|\omega| > 0.05$. This Cartesian grid was organized as a hierarchy of four nested grid levels, and the finest grid was assigned a spatial step size of $dx = D/1024$, where $D$ is the length of the domain. The ratio of the spatial step size on each grid relative to the next coarsest grid was 1:4. The temporal resolution was varied to ensure stability. Each Lagrangian point of the immersed structure was chosen to be $\frac{D}{2048}$ apart (twice the resolution of the finest fluid grid).

%
%

\section{Results}
In this paper, we describe the bulk flow structure within a two-chambered embryonic heart containing both trabeculae and blood cells. The $Wo$ is varied from $0.2$ to $12$, and the trabecular heights are varied from half to twice the biologically relevant case. Note that we consider $Wo$ beyond the biologically relevant range for embryonic zebrafish to gain insight into why hearts may change shape and pumping properties as they grow in developmental or evolutionary time. We consider trabeculae heights outside of the biologically relevant range to gain insights into whether or not physical factors constrain the developing heart to this region of the morphospace.

Streamlines and vorticity plots are used to show the direction of flow and mixing within the heart. We are interested in the direction of flow since endothelial cells are known to sense and respond to not only the magnitude of flow but also to its direction \cite{Wang:2013,Heuslein:2014}. We are also interested in the direction of flow near the cardiac wall since it may alter the advection of morphogens or other signaling agents \cite{Turing:1952,Dan:2005,Wartlick:2009,Howard:2011}. The streamline and vorticity graphs were generated using VisIt visualization software \cite{HPV:VisIt}. When interpreting streamlines, please note that a neutrally buoyant, small particle in the fluid will follow the streamline. The streamlines are drawn by making a contour map of the
stream function, since the stream function is constant along the streamline. The stream function, $\psi({\bf{x}}, t)$, in 2D is defined by
the following equations:
\begin{eqnarray}
u({\bf{x}},t)=\frac{\partial \psi({\bf{x}},t)}{\partial y}\\
v({\bf{x}},t)=-\frac{\partial \psi({\bf{x}},t)}{\partial x}
\end{eqnarray}

The vorticity, ${\bf{\omega}}$, is the curl of the velocity field and describes the \emph{local} rotation of the fluid. 
\begin{equation}
\label{vorticity} {\bf{\omega}} = \nabla \times {\bf{u}}.
\end{equation}

\begin{figure}
\centering
\includegraphics[width=6.5in]{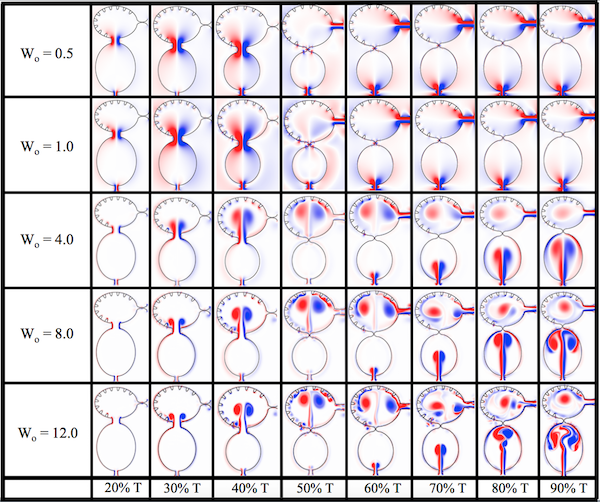}
\caption{Vorticity analysis performed for the case of biologically sized trabeculae and varying $Wo$ at different time points during one heart cycle.}
\label{Wo_Vorticity_Sweep}
\end{figure}

\begin{figure}[H]
    \centering
    \begin{subfigure}{0.9\textwidth}
        \centering
        \includegraphics[width=\textwidth]{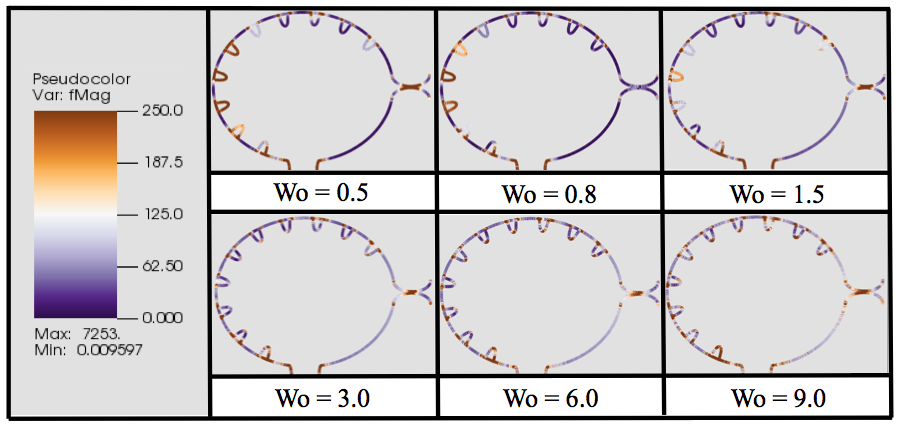}
        \caption{}
        \label{fMag_Wo_Sweep}
    \end{subfigure}\\
    \begin{subfigure}{0.9\textwidth}
        \centering
        \includegraphics[width=\textwidth]{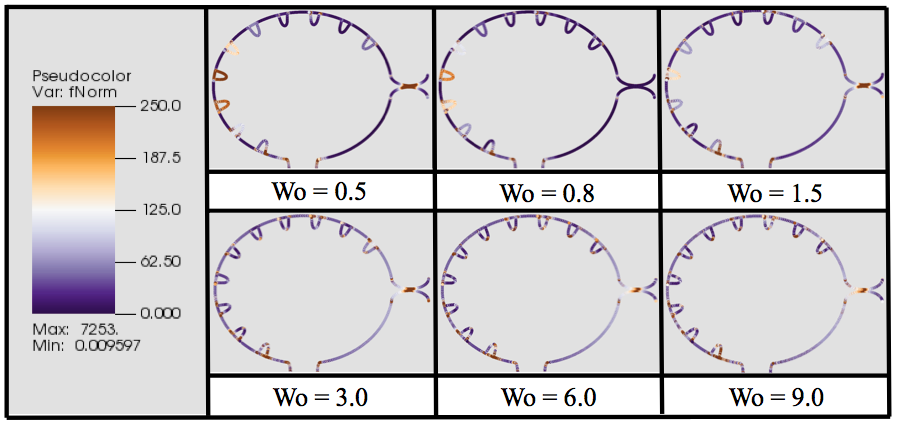}
        \caption{}
        \label{fNorm_Wo_Sweep}
    \end{subfigure}\\ 
    \begin{subfigure}{0.9\textwidth}
        \centering
        \includegraphics[width=\textwidth]{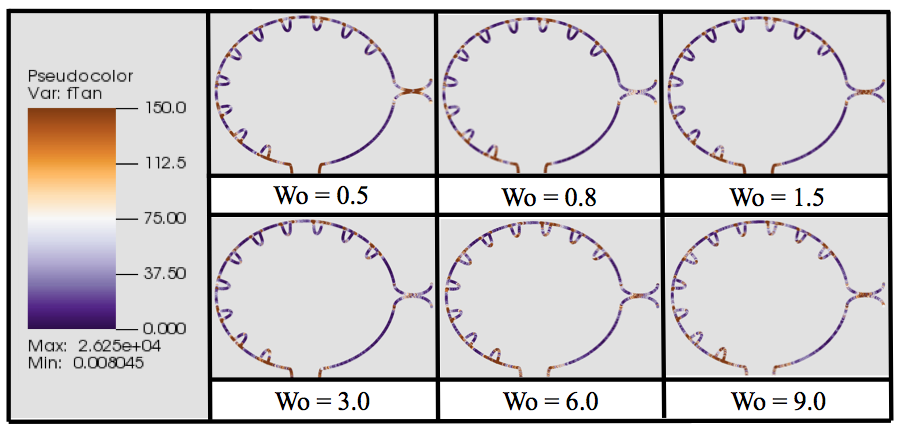}
        \caption{}
        \label{fTan_Wo_Sweep}
    \end{subfigure}\\ 
\caption{}
\label{force_Wo_Sweep}
\end{figure}

Figure \ref{Wo_Vorticity_Sweep} shows the vorticity within the two-chambered heart at different times during one period of the heart cycle, $T$. The trabeculae heights were fixed at the biological scale and no blood cells were simulated. Five different $Wo$ were considered, $Wo=\{0.5,1.0,4.0,8.0,12.0\}$. Note that the biologically relevant case is $Wo=0.77$, which falls between the $Wo=0.5$ and $Wo=1.0$ cases. From the vorticity plots, it is clear there is not much difference in vortical flow between these cases, either during disatole or systole. Furthermore, vortices do not form within the atria during atrial filling. As $Wo$ increases to $Wo=4.0$, two distinct intracardial vortices form, and after systole, a remnant vortex is still present in the ventricle. Two vortices form within the atria during filling. The higher $Wo>4$ cases, show similar vortex existence; however, the ventricular and atrial vortices that form during diastole and systole, respectively, move within the chamber. Moreover, in the $Wo=12.0$ case, distinct vortices are observed between trabeculae, and some minor vortex shedding appears as high speed flow moves over the trabeculae. It is clear as $Wo$ increases, intracardial and intertrabecular mixing also increases. The $Wo$ of adult zebrafish and larger vertebrates is above 4 \cite{Santhanakrishnan:2011}, and this suggest that the role of the trabeculae in the adult may be different than it is during development. It is also interesting to note that adult hearts across the animal kingdom operating at $Wo<4$ typically lack trabeculae.

Figure \ref{force_Wo_Sweep} illustrates the total force magnitude (\ref{fMag_Wo_Sweep}), the normal force magnitude on the boundary(\ref{fNorm_Wo_Sweep}), and the tangential force magnitude on the boundary (\ref{fTan_Wo_Sweep}) for various $Wo$ between $0.5$ and $9.0$ immediately after diastole. It appears the main contribution to the total force magnitude comes from the normal component of the force in all cases. Furthermore it appears that as $Wo$ increases, the force felt on the left most trabeculae decreases, and there is an increase in the forces felt by the top surface of each trabeculae as well as between the trabeculae, i.e., both normal and tangential components of the force, see Figures \ref{fNorm_Wo_Sweep} and \ref{fTan_Wo_Sweep}. 

It is clear from Figure \ref{force_Wo_Sweep} that the region experiences the largest forces is on the left side of the ventricle, e.g., the side opposite to the bulbus arteriosus, as diastole finishes. Next we examined the the average magnitude of the force over a trabeculae over one heart cycle for $Wo=0.8$, see Figure \ref{fMag_bio_time}. During one heart beat, there appears to be three local extrema in the magnitude of the force for trabeculae $\#3,\#4,\#5$, two local maxima and one local minimum. To decipher what forces were dominant, we computed the average magnitude of the normal and tangential components of the force on the trabeculae, as illustrated in Figure \ref{fTanfNorm_compare}. The analysis shows that the normal component of the force dominates for trabeculae $\#3,\#4,\#5$. Moreover, two local maxima and one local minimum are observed for the normal component of the force, averaged over the heartbeat. One local maximum appears for the tangential component of the force for $Wo=0.8$. 

Furthermore, for trabeculae $\#3$, a scaling study was performed for $Wo$ ranging from $0.5$ to $12.0$. Figure \ref{fMag_Wo_Sweep_time} depicts the average magnitude of the force over one heartbeat cycle for trabeculae $\#3$. The analysis yielded similar functional behavior, compared with the analogous case in Figure \ref{fMag_bio_time}, for $Wo\leq3$. For $Wo> 3$, there is a clear bifurcation where the average force magnitude no longer displays three local extrema, but instead appears to monotonically increase to an asymptotic maximum over one heartcycle.  


\begin{figure}[H]
    \centering
    \begin{subfigure}{0.39\textwidth}
        \centering
        \includegraphics[width=\textwidth]{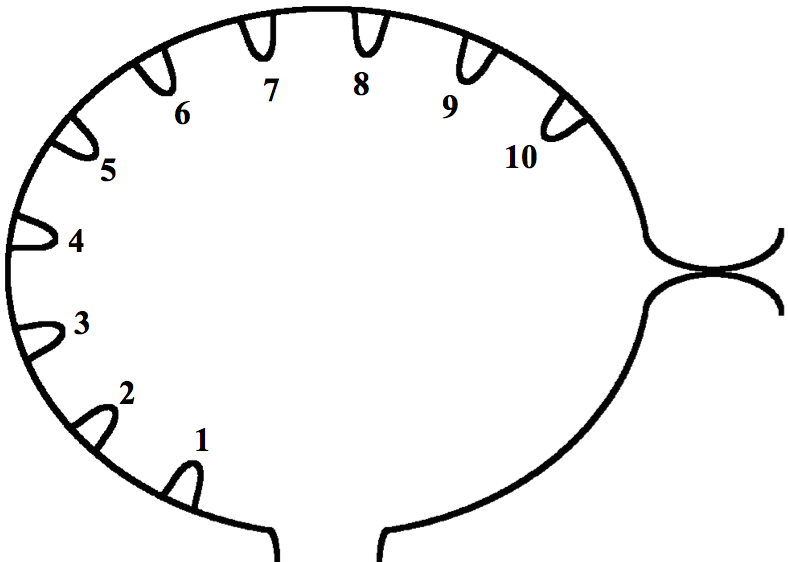}
        \caption{}
        \label{ventricle_geo}
    \end{subfigure}
    \begin{subfigure}{0.6\textwidth}
        \centering
        \includegraphics[width=\textwidth]{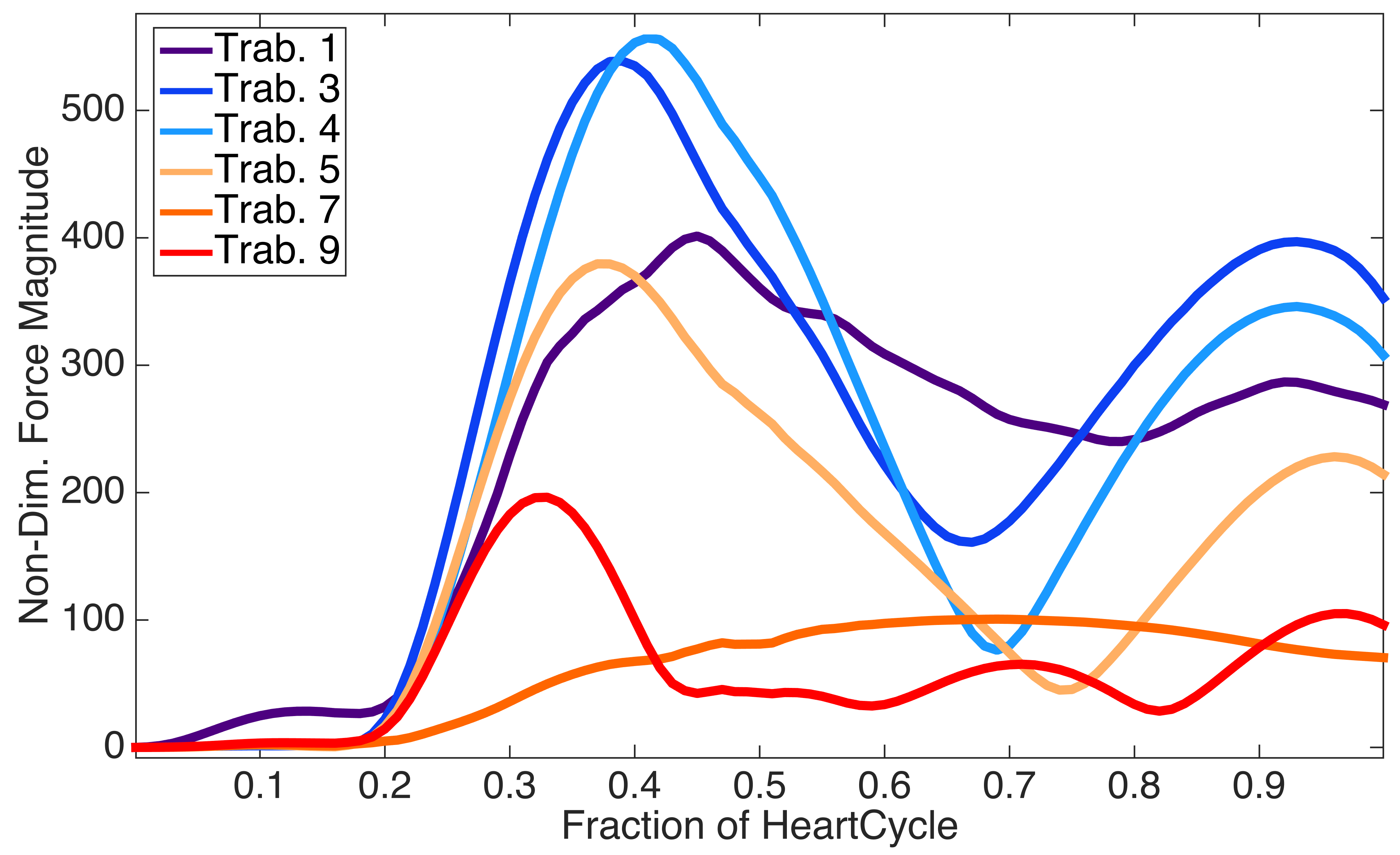}
        \caption{}
        \label{fMag_bio_time}
    \end{subfigure}\\
    \begin{subfigure}{0.47\textwidth}
        \centering
        \includegraphics[width=\textwidth]{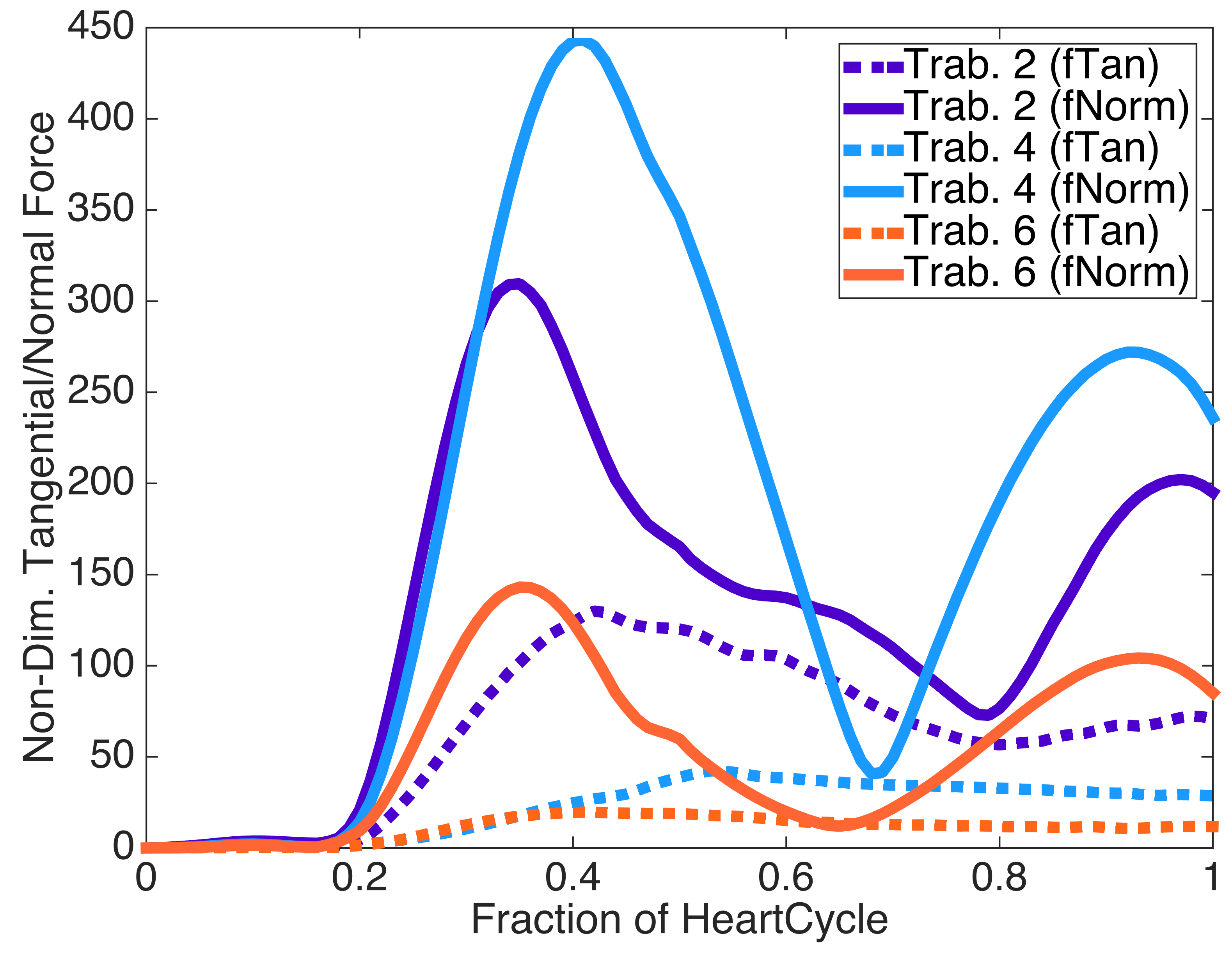}
        \caption{}
        \label{fTanfNorm_compare}
    \end{subfigure}
    \begin{subfigure}{0.51\textwidth}
        \centering
        \includegraphics[width=\textwidth]{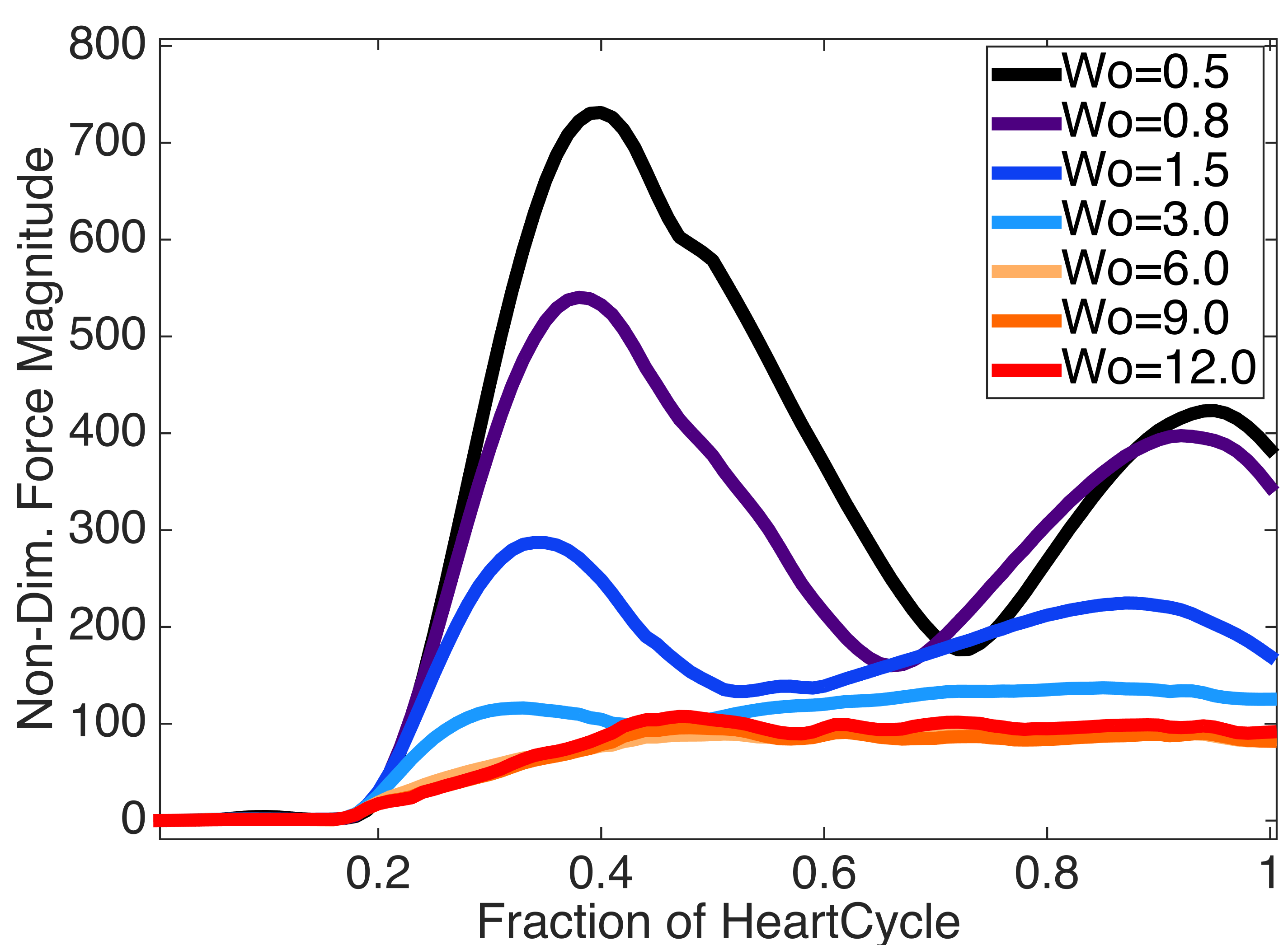}
        \caption{}
        \label{fMag_Wo_Sweep_time}
    \end{subfigure}\\ 
\caption{(a) Illustrating the indexing of trabeculae (b) Plot illustrating the average magnitude of the force on chosen trabeculae over the course of one heart cycle for $Wo=0.8$, the biologically relevant case. (c) Plot showing the average magnitude of the tangential and normal forces at each time, for chosen trabeculae, during one heart cycle for $Wo=0.8$. (d) A plot illustrating the average magnitude of force at each time-step for $Wo$ ranging from $0.5$ (half the biologically relevant case) to $12.0$.}
\label{force_Wo_Sweep}
\end{figure}

%
%

\subsection{Effects of Trabeculae Height}

\begin{figure}[H]
\centering
\includegraphics[width=6.5in]{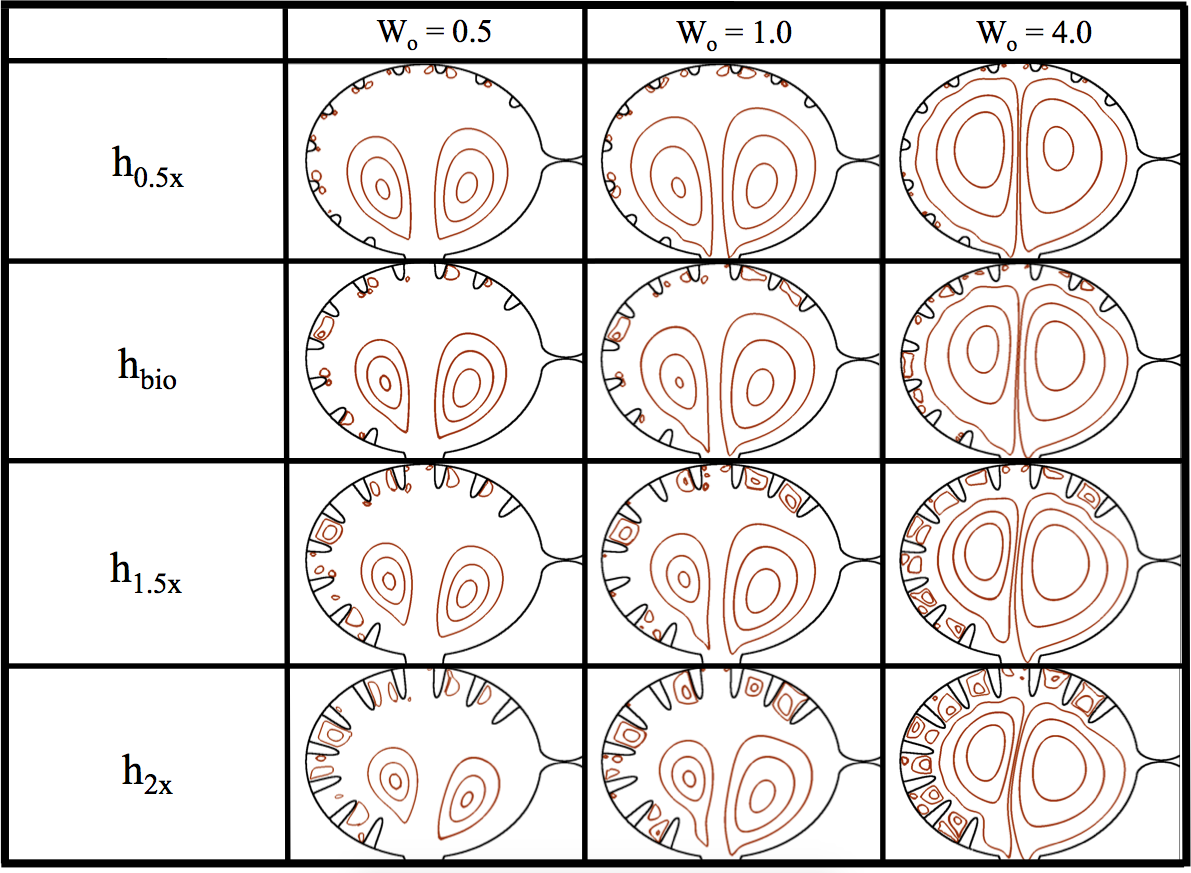}
\caption{Streamline analysis for $Wo=\{0.5,1.0,4.0\}$ and trabecular heights from half to twice the biologically relevant size. The analysis was performed within the ventricle immediately after diastole finishes and when the ventricle stops expanding.}
\label{Wo_Height_Sweep}
\end{figure}

Figure \ref{Wo_Height_Sweep} shows closed streamlines for $Wo=\{0.5,1.0,4.0\}$ and trabecular heights ranging from half to twice the biologically relevant size. No blood cells were added to the simulations, and the trabeculae radii and locations were fixed, keeping them equidistant along the ventricular chamber. The analysis was performed within the ventricle immediately after diastole finishes, e.g., the ventricle stops expanding.

\begin{figure}[H]
\centering
\includegraphics[width=6.5in]{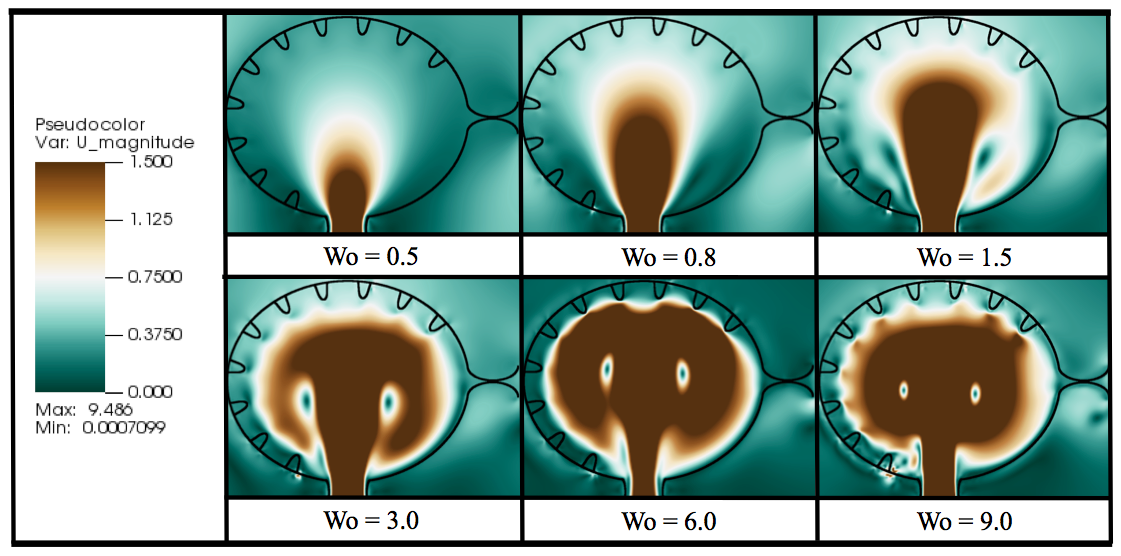}
\caption{Magnitude of velocity colormaps, corresponding to simulations of varying $Wo$ for biologically relevant trabeculae height. The images were taken immediately after diastole, when the ventricle stops expanding.}
\label{uMag_Plots}
\end{figure}

In the $h_{0.5x}$ case, i.e., half the biologically relevant height, some small vortices appear to form between trabeculae, as seen by the closed streamlines. These closed loops are small relative to the intertrabecular spacing. Note also that the flow velocities between the trabeculae are also quite small, see Figure \ref{uMag_Plots}, which illustrates the magnitude of velocity for simulations of varying $Wo$ for biologically relevant trabeculae heights. Large intracardial vortices are clearly present in all cases, and the size and strength of these vortices grow as $Wo$ increases.  

As the trabeculae height increases, the intertrabecular vortices grow larger. The intracardial vortices remain approximately the same size as height increases. Note also that the intracardial vortex pair becomes more asymmetric as the trabeculae increase in height. Moreover, in the $h_{1.5x}$ and $h_{2x}$ cases, as $Wo$ increases intertrabecular vortices become larger and increase in number (note that these regions still represented relatively slow flow). 

The intracardial vortices both spin in opposite directions, e.g., the vortex to the left rotates counter-clockwise while the vortex on the right spins clockwise. Therefore, the intertrabecular vortices on the left side of the ventricle, which form near the head of the trabeculae, spin clockwise, and vice versa on the opposite side of the ventricle. Furthermore, there is a somewhat stagnant region opposite to the AV canal, where the two vortices diverge, and hence no large intertrabecular vortices form, as compared to different intertrabecular regions in the same simulation. There are small vortices that form in the intertrabecular region opposite to the AV canal in the $h_{0.5x}$ cases; however, as the trabeculae increase in height in the $Wo=0.5$ and $Wo=1.0$ cases, this region becomes scarce of vortical flow, while there remains a small amount in the $Wo=4.0$ case.

\begin{figure}[H]
\centering
\includegraphics[width=6.5in]{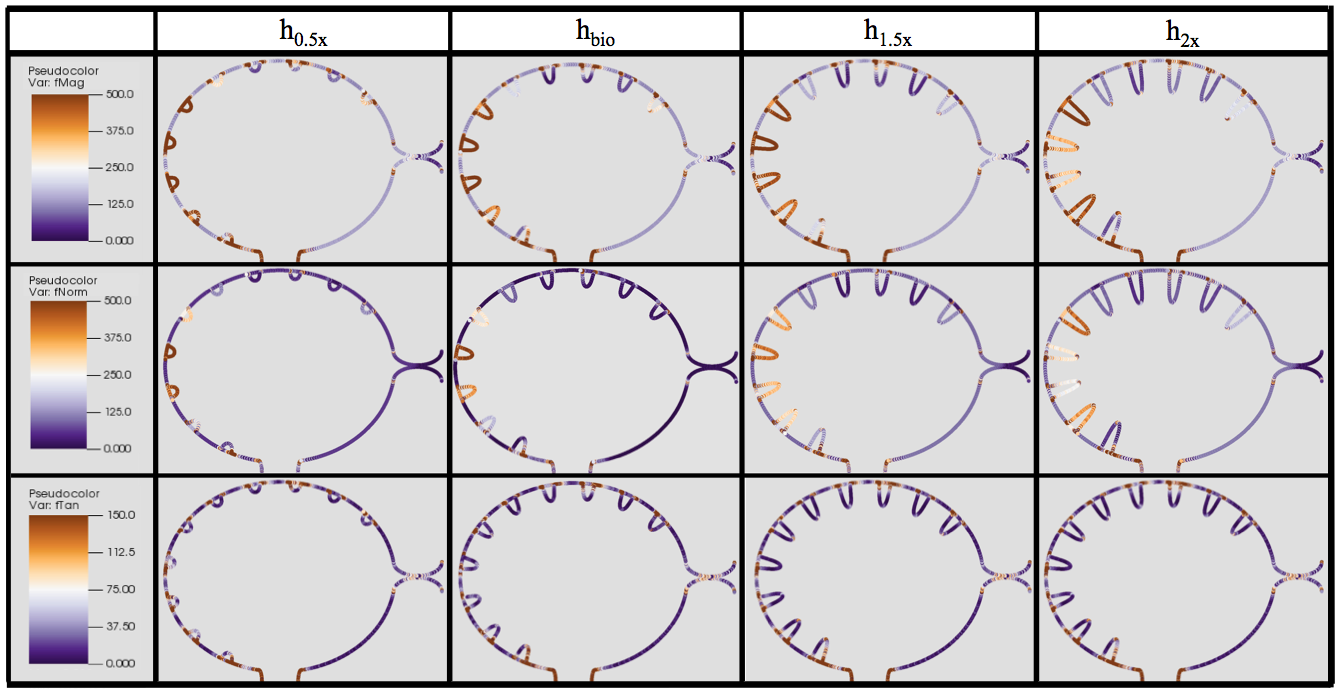}
\caption{The total magnitude of force (top row), magnitude of the normal force to the boundary (middle row), and magnitude of the tangential force to the boundary (bottom) for different trabeculae heights at the biologically relevant $Wo$, $Wo=0.8$. It is clear that while the tangential and normal force magnitudes differ, the main contributor to total force on the boundary is the normal component. The largest force are felt by the trabeculae on the left most side of the ventricle. As the trabeculae height increases, the most force is still felt on the same trabeculae, but with reduced magnitude.}
\label{Force_Height_Sweep}
\end{figure}

For biologically relevant $Wo=0.8$, in the case of $h_{0.5x}$, the most force is felt by the trabeculae on the left side of the ventricle, see Figure \ref{Force_Height_Sweep}. This trend continues regardless of height; however, as the height increases the magnitude of that force is decreased. Furthermore, as the height increases, the trabeculae that are exposed to the most force are no longer the left most trabeculae, but the trabeculae next to them. In all cases, the main contribution to the total force magnitude comes from the normal component of the force, rather than the tangential component.

%
%

\subsection{Effects of Hematocrit}

\begin{figure}[H]
\centering
\includegraphics[width=6.5in]{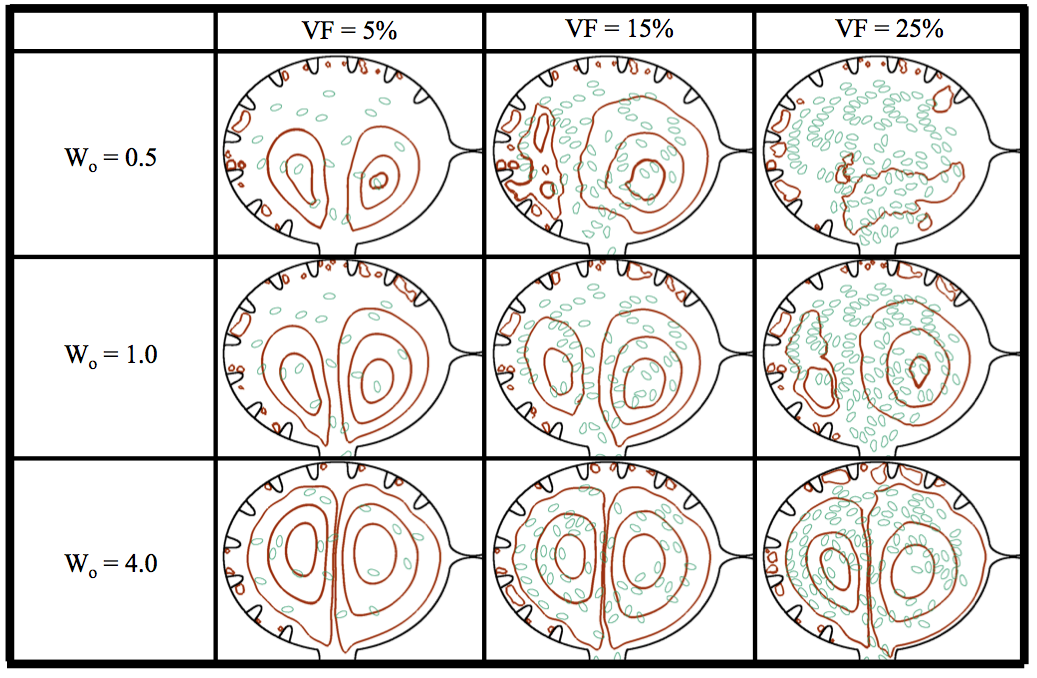}
\caption{Streamline analysis for $Wo=\{0.5,1.0,4.0\}$ and hematocrit of $VF= \{5\%,15\%,25\%\}$. The analysis was performed within the ventricle immediately after diastole finishes and the ventricle ceases its expansion.}
\label{Wo_VF_Sweep}
\end{figure}

Figure \ref{Wo_VF_Sweep} shows the effect that the addition of blood cells has in flow patterns over a range of $Wo$. In these simulations, trabeculae height, radii, and spacing were fixed. Trabeculae heights were modeled at the biologically relevant size. The analysis was performed within the ventricle immediately at the completion of diastole.  

In the case of $Wo=0.5$, it is clear the addition of blood cells alters the flow pattern within the ventricle. When $VF=5\%$, the flow resembles that of the analogous case with no hematocrit as seen in Figure \ref{Wo_Height_Sweep}; however, as hematocrit is increased to $VF=15\%$, the flow patterns are very different. For $VF=5\%, 15\%$, there are still coherent right or clockwise rotating vortices (shown as the closed streamlines) that form on the right side of the ventricle. The right vortex stretches directly above the AV canal and the left vortex is reduced. For $VF = 25\%$, coherent intracardial vortices are not evident. For larger $Wo$ ($Wo=1, 4)$, the coherent intracardial vortex pair is observed even for higher hematocrit. 


In general as $Wo$ increases, the intracardial vortices become more well-defined, e.g., vortical flow is smoother. However, in all cases hematocrit still affects vortical flow patterns intracardially as well as intertrabeularly. Moreover, in these simulations it is clear that after diastole, no blood cells have moved between trabeculae, but rather stay within the middle of the chamber, regardless of volume fraction of $Wo$. These results suggest that for larger adult vertebrates, when the relative size of the blood cells would also be smaller, the presence of blood cells does not dramatically change the bulk flow. The blood cells do appear to affect the formation of coherent intracardial vortices at this stage of development when $Wo = 0.5-1$.

\begin{figure}[H]
\centering
\includegraphics[width=6.5in]{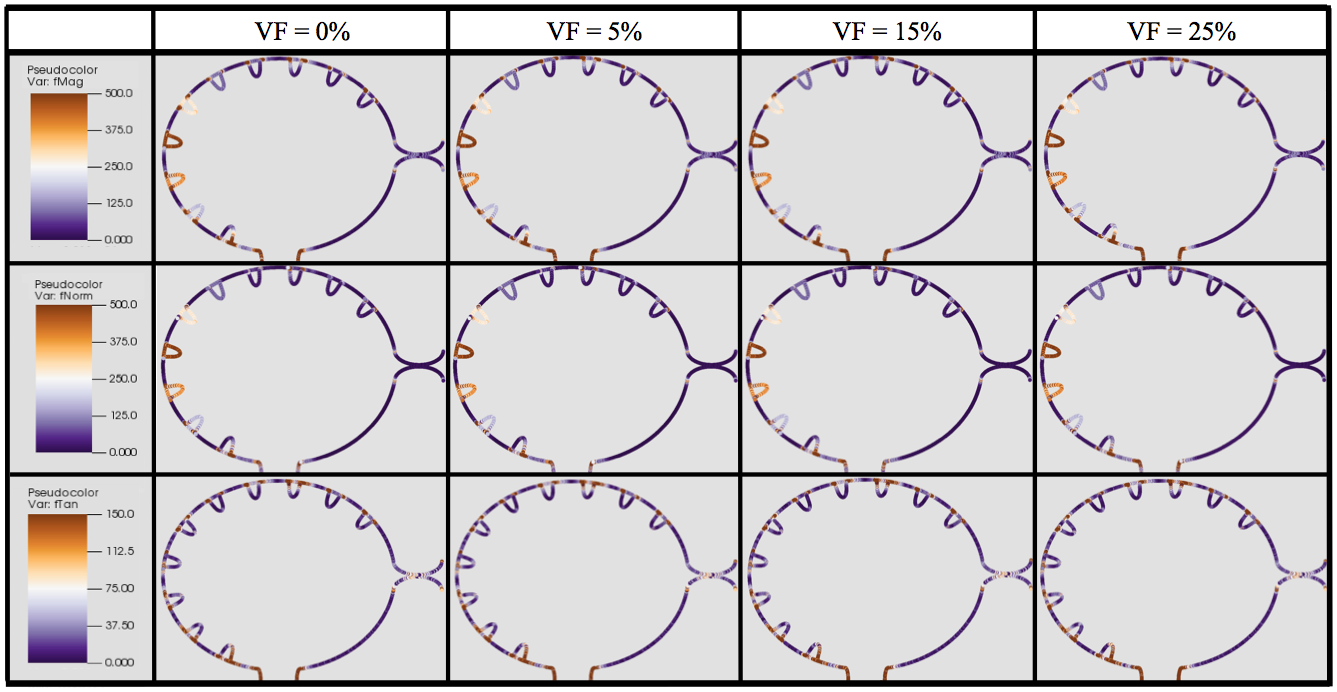}
\caption{The total magnitude of force (top row), magnitude of the normal force to the boundary (middle row), and magnitude of the tangential force to the boundary (bottom) for different volume fractions at the biologically relevant $Wo$, $Wo=0.8$. It is clear that while the tangential and normal force magnitudes differ, the main contributor to total force on the boundary is the normal component. Moreover, the blood cells do not appear to affect the magnitude of the force on the boundary in any case.}
\label{Force_VF_Sweep}
\end{figure}

Although the blood cells do affect intracardial vortices, they do not appear to have significant affect on the magnitude of forces on the boundary for biologically relevant $Wo=0.8$, see Figure \ref{Force_VF_Sweep}. It is clear that the main contribution to the total force comes from the normal force component to the boundary, but the net differences between the cases for each volume fraction are minute. This may be the case due to the blood cells having not been thrust close enough to the intertrabecular regions.

%
%

\subsection{Fluid Mixing}

\begin{figure}[H]
    \begin{subfigure}{0.5\textwidth}
        \centering
        \includegraphics[width=\textwidth]{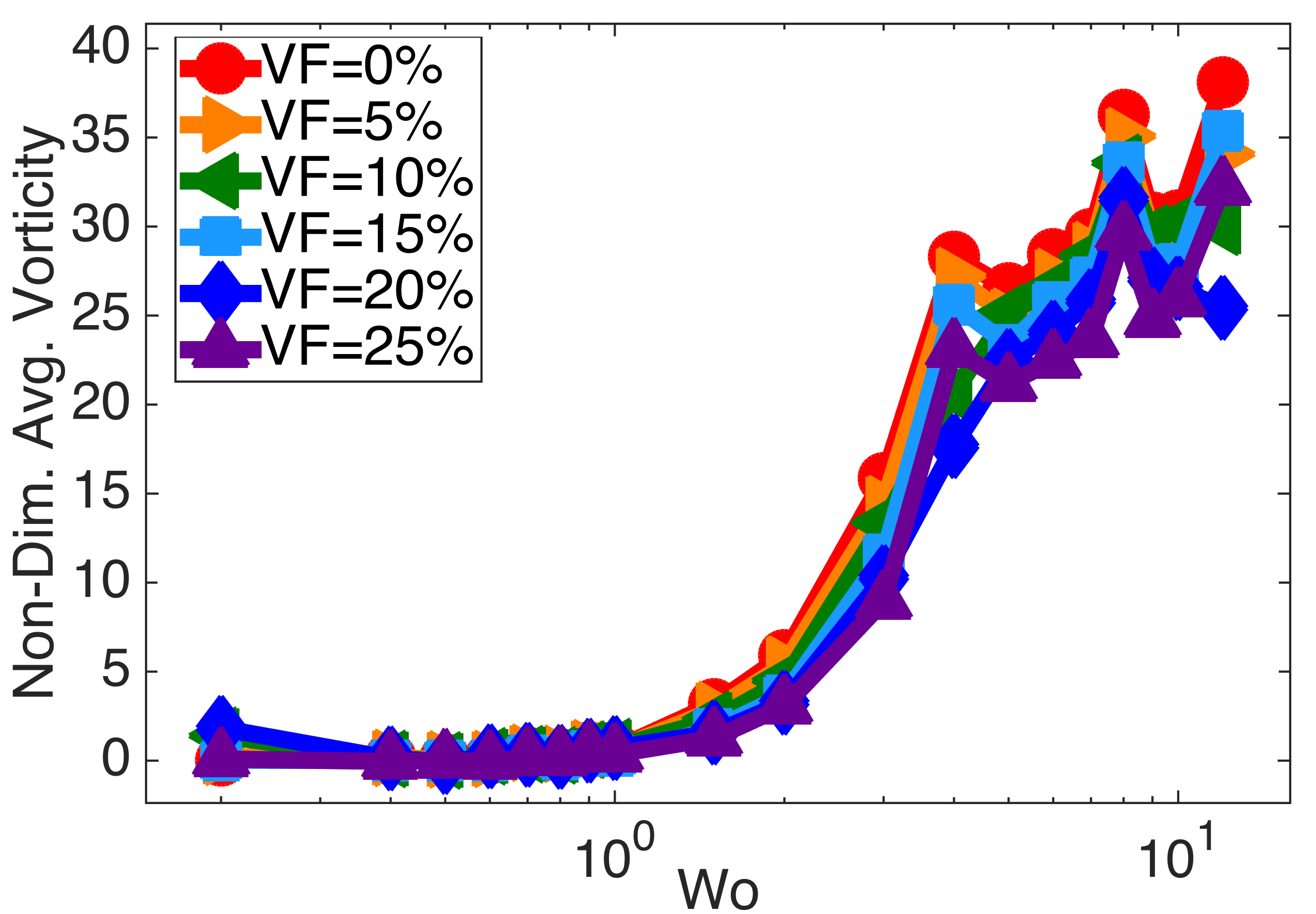}
        \caption{}
        \label{VorticityLeft}
    \end{subfigure}
    \begin{subfigure}{0.5\textwidth}
        \centering
        \includegraphics[width=\textwidth]{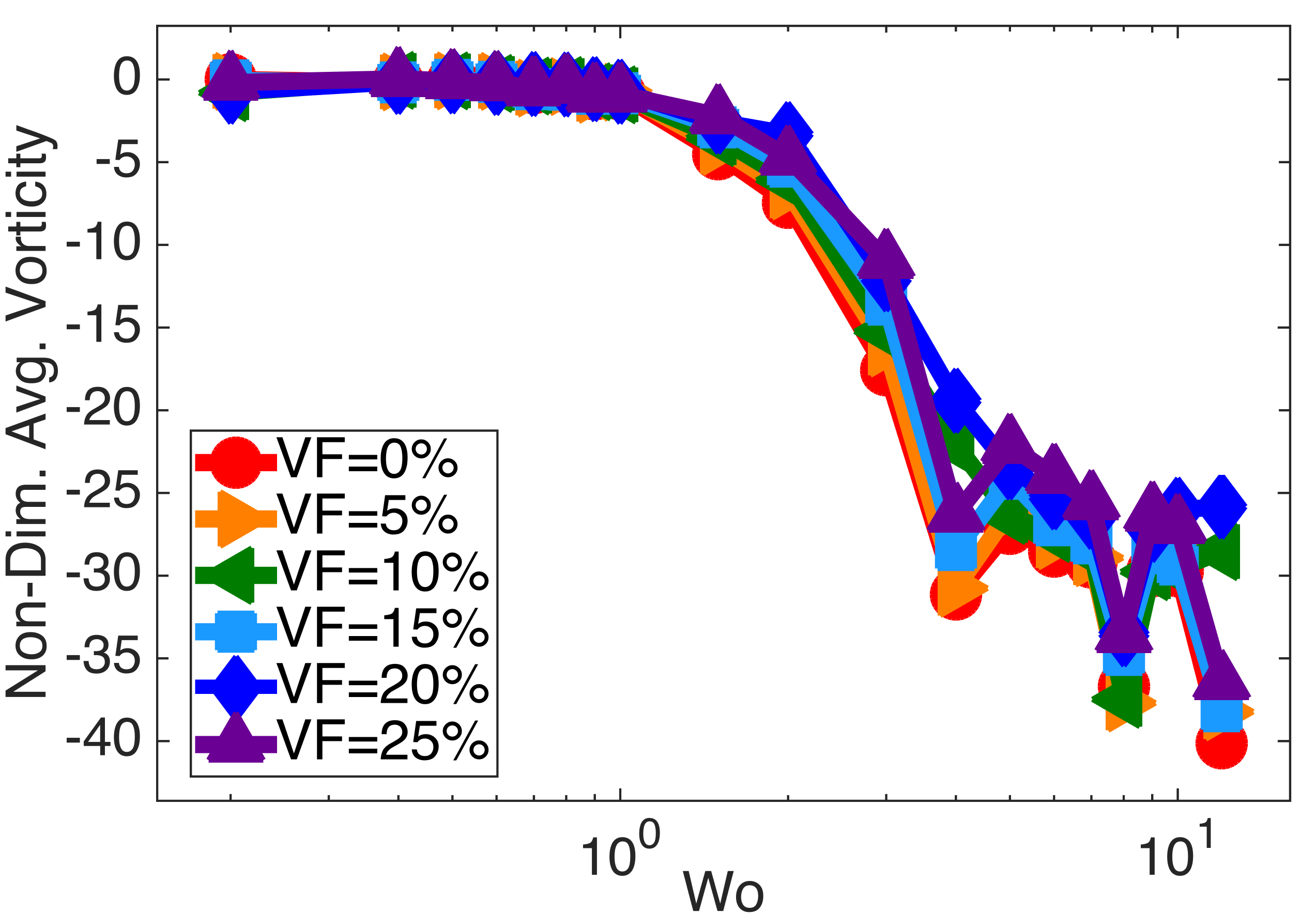}
        \caption{}
        \label{VorticityRight}
    \end{subfigure}\\ \\
\caption{\ref{VorticityLeft} and \ref{VorticityRight} illustrate the average fluid vorticity on the left and right side of the ventricle, respectively, immediately after diastole, as a function of $Wo$. It is clear there is a non-linear relationship between the spatially-averaged vorticity and biological scale, given by $Wo$.}
\label{VorticityVsWo}
\end{figure}

As a rough approximation of the rotation and mixing in the fluid, we calculated that spatially-averaged vorticity in the ventricle. Figures \ref{VorticityLeft} and \ref{VorticityRight} give the spatially-averaged fluid vorticity on the left and right side of the ventricle, respectively, immediately after diastole, as a function of $Wo$ for  $VF=\{0\%,5\%,\ldots,25\%\}$. It is evident that there is a non-linear relationship between spatially-averaged vorticity and $Wo$. Furthermore, the overall net sign of the spatially-averaged vorticity is positive in the left side of the ventricle, while it is opposite on the right side. Moreover, the presence of blood cells does not appear to significantly affect the spatially-averaged fluid vorticity for $Wo\leq1$, although it does affect the generation of a coherent vortex pair.  

We report the spatially-averaged vorticity at different times during an entire heartbeat in each side of the ventricle in Figure \ref{VorticityHeartCycle}. The spatially averaged vorticity was calculated for $Wo=\{0.5,0.8,1.5,8.0\}$ (note the biologically relevant case is $Wo=0.8$) for hematocrit, $VF=\{0\%,5\%,15\%,25\%\}$. When $Wo\leq 1.5$, there is a clear peak before diastole ends (the vertical dotted line), while for $Wo=8.0$, the peak occurs the moment when diastole ends. Note also that the width of this peak is larger for $Wo=8.0$ when inertia dominates. 

In general as hematocrit increases, so does the spatially-averaged vorticity. Locally, the presence of blood cells act to increase vorticity in either direction through their tumbling motion, and this enhancement is not captured in the spatial average.


\begin{figure}[H]
    \begin{subfigure}{0.5\textwidth}
        \centering
        \includegraphics[width=0.85\textwidth]{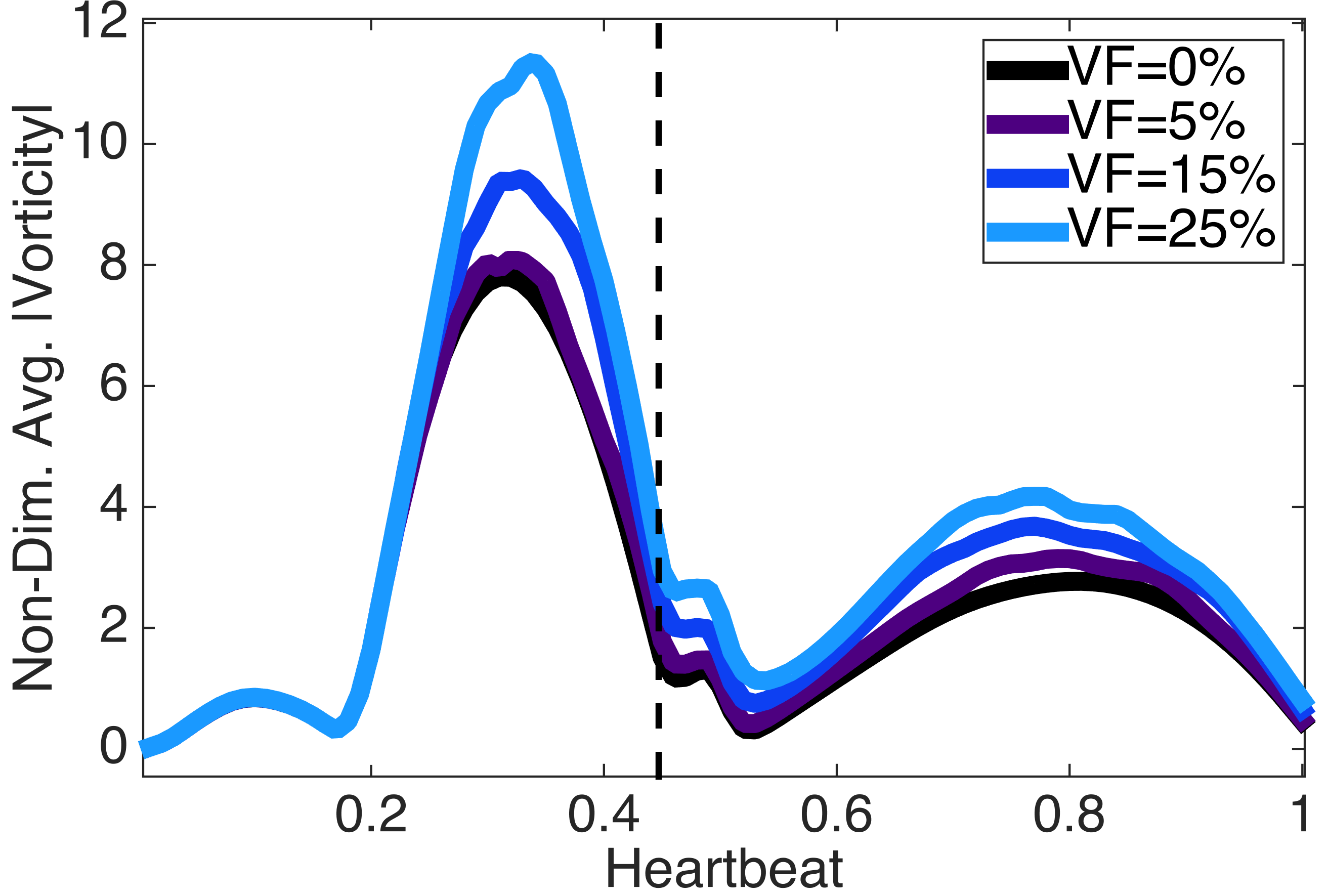}
        \caption{$Wo=0.5$}
        \label{V_Wo0pt5}
    \end{subfigure}
    \begin{subfigure}{0.5\textwidth}
        \centering
        \includegraphics[width=0.85\textwidth]{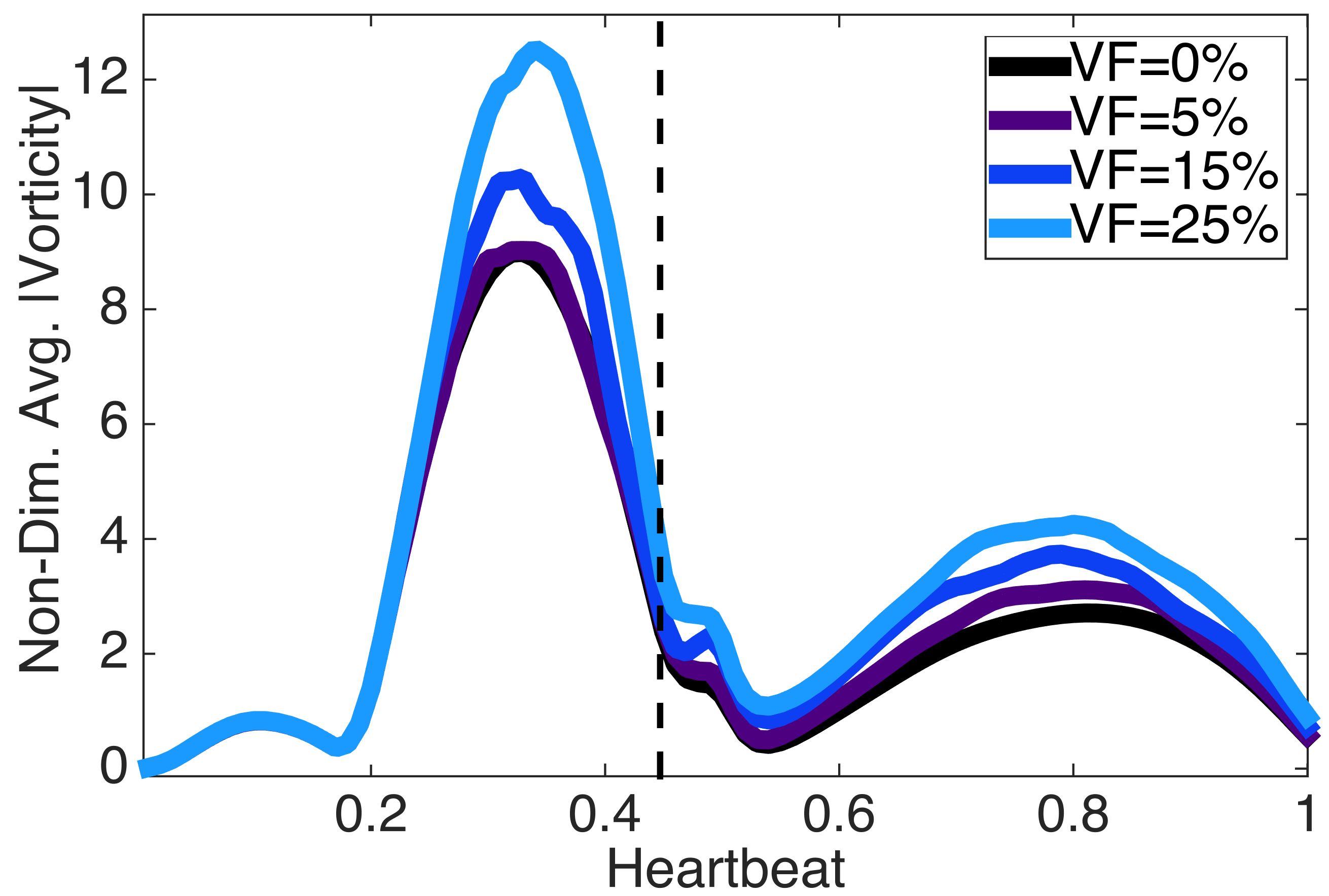}
        \caption{$Wo=0.8$}
        \label{V_Wo0pt8}
    \end{subfigure}\\ \\
        \begin{subfigure}{0.5\textwidth}
        \centering
        \includegraphics[width=0.85\textwidth]{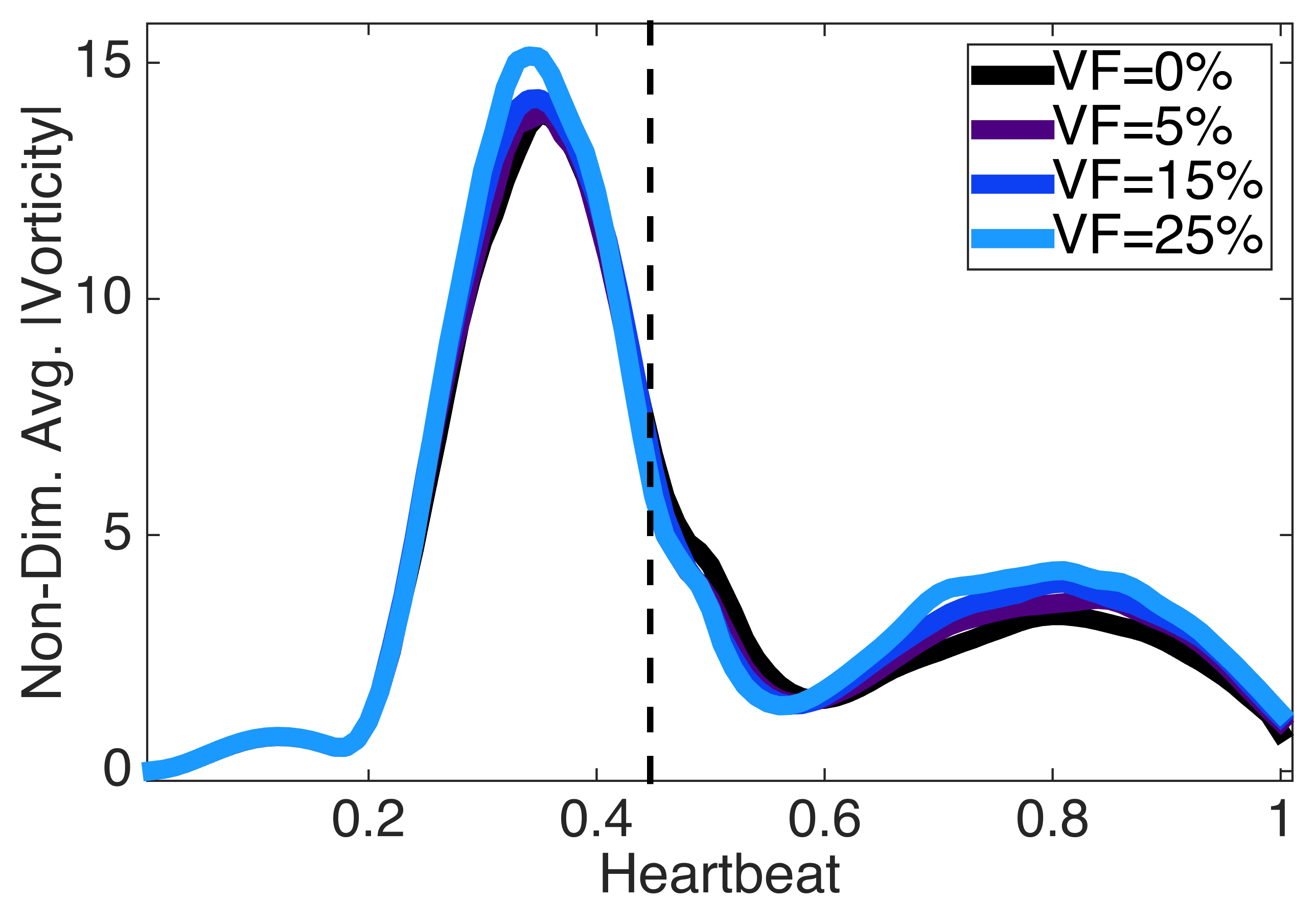}
        \caption{$Wo=1.5$}
        \label{V_Wo1pt5}
    \end{subfigure}
    \begin{subfigure}{0.5\textwidth}
        \centering
        \includegraphics[width=0.85\textwidth]{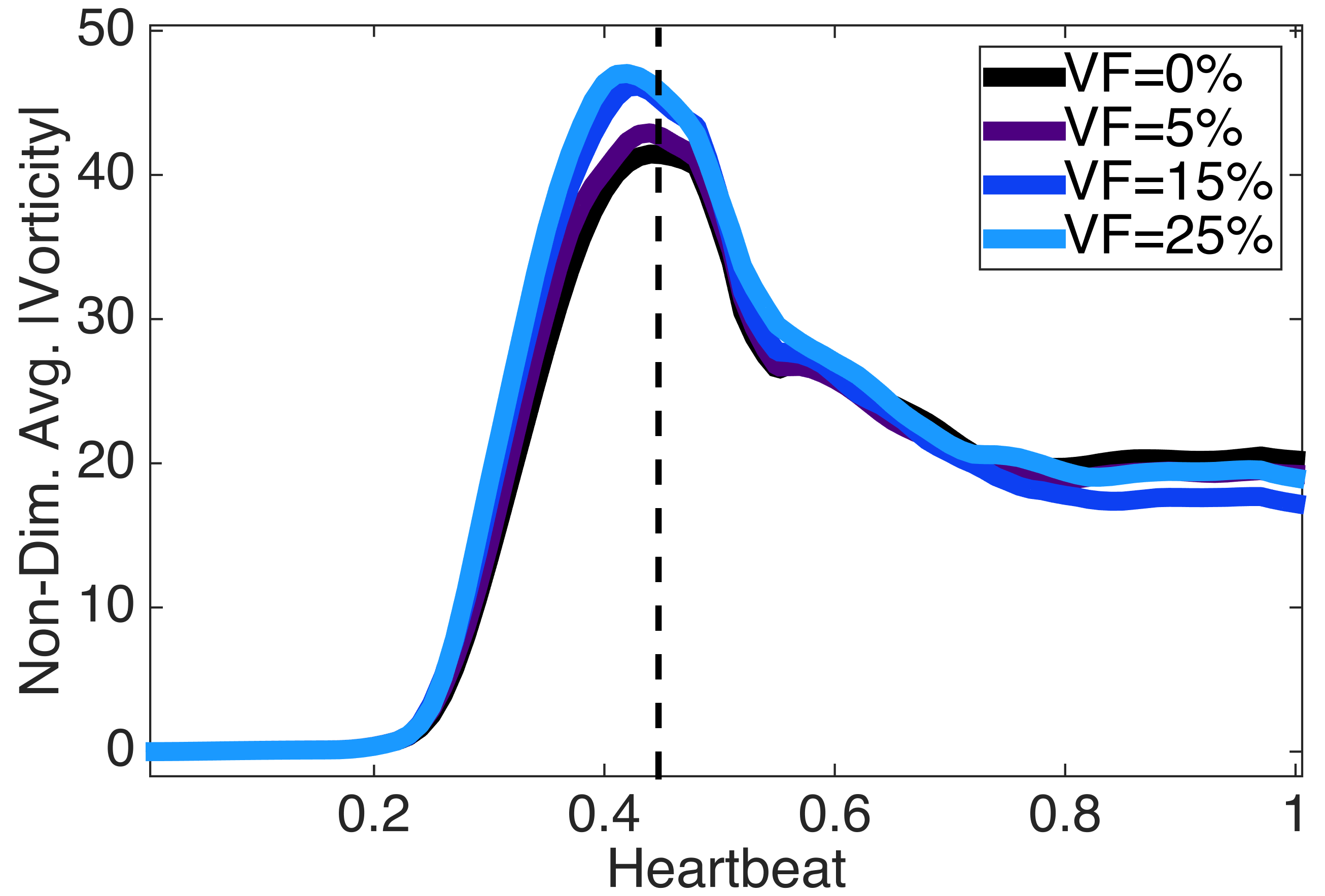}
        \caption{$Wo=8.0$}
        \label{V_Wo8}
    \end{subfigure}\\ \\
\caption{Plots of the spatially-averaged magnitude of vorticity for $Wo=\{0.5,0.8,1.5,8.0\}$ for $VF=\{0\%,5\%,15\%,25\%\}$. The vertical dotted line indicates when diastole ends. For every case of $Wo$, the higher the hematocrit, the more spatially-averaged vorticity magnitude is induced.}
\label{VorticityHeartCycle}
\end{figure}

%
%

\section{Conclusions}

Two-dimensional immersed boundary simulations were used to solve for the fluid motion within an idealized two-chambered pumping heart. The presence of blood cells, trabeculae, and the relative importance of unsteady effects (e.g. the $Wo$) were considered.  The geometry models an idealized embryonic zebrafish heart at $4\ dpf$, and the motion of the chambers was approximated from the kinematic analysis of video taken from a wild type embryonic zebrafish. The main results of the study are as follows: 1) without the presence of blood cells, a large vortex pair forms in the ventricle during filling; 2) with the presence of blood cells at lower $Wo$, a coherent vortex pair is not formed, 3) for $Wo>4$, intertrabecular vortices form and vorticity separates from the trabeculae (suggesting the effect of the trabeculae is different in adult vertebrates than in embryos); 4) the presence of blood cells enhances spatially averaged vorticity in the ventricle, which peaks during diastole; 5) the presence of blood cells does not significantly alter the forces felt by the endocardial cells, and 6) the majority of force is felt by the trabeculae on the outer region of the ventricle. 

As mentioned above, an oppositely spinning large intracardial vortex pair forms for all $Wo$ considered, here for $Wo>0.2$. The vortex on the left spins counterclockwise, while the vortex on the right spins clockwise. This distinction becomes important when considering the formation of vortices between trabeculae.  Larger intertrabecular vortices form for simulations with taller trabeculae. Furthermore, when the trabeculae height was $1.5x$ or $2x$ the biologically relevant height in the $Wo=4$ case, stacked vortices formed between trabeculae; the top vortex spinning opposite to that of the closest intracardial vortex, while the vortex near the base spinning opposite to that. With the addition of blood cells, coherent intracardial vortices do not form when $Wo<4$ and $VF\leq15\%$; however, intertrabecular vortical flow patterns were not significantly changed as blood cells were not advected into these regions. 

Note that the presence or absence of vortices alter the magnitude and direction of flow near the endocardial wall as well as the mixing patterns within the ventricle. When an intracardial vortex forms, the direction of the flow changes. The presence of two large intracardial vortices forms a stagnation point on the opposite side of the ventricle to the AV canal. Also, the presence of intertrabecular vortices changes the direction of the flow between trabeculae; not all intertrabecular regions have the formation of these vortices. In such cases, the direction of flow between different trabeculae will move in different directions. Since endothelial cells are known to sense and respond to changes in both magnitude and direction of flow \cite{Heuslein:2014}, the formation and motion of intracardial and intertrabecular vortices may be important epigenetic signals. 

For the biologically relevant parameter choices, $Wo$ between $0.5$ and $1.0$, it is clear that the addition of blood cells significantly affects the formation of coherent vortices. This illustrates the importance of considering hematocrit when conducting fluid dynamics studies at this stage of development. Furthermore, this study demonstrates that small changes in viscosity, scale, morphology, and hematocrit can influence bulk flow properties in the embryonic heart. This presents an interesting challenge since each of these parameters are continuously changing during heart morphogenesis. In addition, estimating the effective viscosity and hematocrit of the embryonic blood is nontrivial. 

The results of this paper demonstrate the importance of scale, morphology, and the presence of blood cells in determining the bulk flow patterns through the developing heart. This is important because there is a strongly coupled relationship between intracardial hemodynamics, genetic regulatory networks, and cardiac conduction \cite{Hu:1989,Hove:2003,Bartman:2004,Scherz:2008,Culver:2010,Santhanakrishnan:2011,Garita:2011,Granados:2012,Jamison:2013,Chen:2014,Samsa:2015}. Besides contractions of the myocardial cells, which in turn drive blood flow, hemodynamics are directly involved in proper pacemaker and cardiac conduction tissue formation. \cite{Tucker:1988}. Moreover, shear stress is found to regulate spatially dependent conduction velocities within the myocardium. \cite{Reckova:2003}. Myocardial contractions are also required for trabeculation \cite{Samsa:2015}. It is important to note that changes in the conduction properties of the embryonic heart will also affect the intracardial shear stresses and pressures and patterns of cyclic strains.

The cyclic stresses and strains of the cardiomyocytes can also help shape the overall architecture of the trabeculated ventricle. The dynamics of these strains depend upon the intracardial fluid dynamics. For example, greater resistant to flow will induce larger cyclic stresses and possibly reduced cyclic strains. It is known that cyclic strains initiate myogenesis in the cellular components of primitive trabeculae. \cite{Garita:2011} Since trabeculation first occurs near peak stress sites in the ventricle, altering blood flow may directly produce structural and morphological abnormalities in cardiogenesis. Previous work focusing on hemodynamic unloading in an embryonic heart has resulted in disorganized trabeculation and arrested growth of trabeculae \cite{Sankova:2010,Peshkovsky:2011,Bartman:2004}. On the other hand, embryos with a hypertrabeculated ventricle also experience impaired cardiac function. \cite{Peshkovsky:2011}

The exact mechanisms of mechanotransduction are not yet clearly understood \cite{Weinbaum:2003,Paluch:2015}. Biochemical signals are thought to be propagated throughout a pipeline of epigenetic signaling mechanisms, which may lead to a regulation of gene expression, cellular differentiation, proliferation, and migration \cite{Chen:2014}. \textit{In vitro} studies have discovered that endothelial cells can detect shear stresses as low as $0.2\ dyn/cm^2$ \cite{Olesen:1988} resulting in up or down regulation of gene expressions. Embryonic zebrafish hearts around $36$ hpf are believed to undergo shear forces of $\sim2\ dyn/cm^2$ and shearing of $\sim75\ dyn/cm^2$ by $4.5$ dpf \cite{Hove:2003}. Shear forces in the $\sim8-15\ dyn/cm^2$ range are known to cause cytoskeletal rearrangement \cite{Davies:1986}. Mapping out the connection between fluid dynamics, the resulting forces, and the mechanical regulation of developmental regulatory networks will be critical for a global understanding of the process of heart development.

%
%

\section{Acknowledgements}
The authors would like to thank Steven Vogel for conversations on scaling in various hearts. We would also like to thank Lindsay Waldrop, Austin Baird, Leigh Ann Samsa, and William Kier for discussions on embryonic hearts. This project was funded by NSF DMS CAREER \#1151478, NSF CBET \#1511427, NSF DMS \#1151478, and NSF POLS \#1505061 awarded to L.A.M. Funding for N.A.B. was provided from an National Institutes of Health T32 grant [HL069768-14; PI, Christopher Mack].

%
%

\newpage

\bibliographystyle{spmpsci}      
\bibliography{heart}   

\begin{thebibliography}{10}
\providecommand{\url}[1]{{#1}}
\providecommand{\urlprefix}{URL }
\expandafter\ifx\csname urlstyle\endcsname\relax
  \providecommand{\doi}[1]{DOI~\discretionary{}{}{}#1}\else
  \providecommand{\doi}{DOI~\discretionary{}{}{}\begingroup
  \urlstyle{rm}\Url}\fi

\bibitem{Roubaie:2011}
Al-Roubaie, S., Jahnsen, E.D., Mohammed, M., Henderson-Toth, C., Jones, E.A.:
  Rheology of embryonic avian blood.
\newblock Am. J. Physiol. Heart Circ. Physiol. \textbf{301}(6919), 2473--2481
  (2011)

\bibitem{Bartman:2004}
Bartman, T., Walsh, E.C., Wen, K.K., McKane, M., Ren, J., Alexander, J.,
  Rubenstein, P.A., Stainier, D.Y.: Early myocardial function affects
  endocardial cushion development in zebrafish.
\newblock PLoS Biol \textbf{2}, E129 (2004)

\bibitem{MJBerger84}
Berger, M., Oliger, J.: Adaptive mesh refinement for hyperbolic
  partial-differential equations.
\newblock J~Comput~Phys \textbf{53}(3), 484--512 (1984)

\bibitem{MJBerger89}
Berger, M., P.Colella: Local adaptive mesh refinement for shock hydrodynamics.
\newblock J~Comput~Phys \textbf{82}(1), 64--84 (1989)

\bibitem{Bhalla:2013a}
Bhalla, A., Griffith, B., Patankar, N.: A forced damped oscillation framework
  for undulatory swimming provides new insights into how propulsion arises in
  active and passive swimming.
\newblock PLOS Comput. Biol. \textbf{9}, e1003,097 (2013)

\bibitem{Burggren:2004}
Burggren, W.: What is the purpose of the embryonic heart beat? or how facts can
  ultimately prevail over physiological dogma.
\newblock Physiological and Biochemical Zoology \textbf{77}, 333–345 (2004)

\bibitem{Cartwright:09}
Cartwright, J.H., Piro, O., Tuval, I.: Fluid dynamics in developmental biology:
  moving fluids that shape ontogeny.
\newblock HFSP J. \textbf{3}, 77–93 (2009)

\bibitem{Chapman:1918}
Chapman, W.: The effect of the heart-beat upon the development of the vascular
  system in the chick.
\newblock Am. J. Anat. \textbf{23}, 175–203 (1918)

\bibitem{Chen:2014}
Chen, L., Wei, S., Chiu, J.: Mechanical regulation of epigenetics in vascular
  biology and pathobiology.
\newblock J. Cell. Mol. Med. \textbf{11}, 437--448 (2014)

\bibitem{HPV:VisIt}
Childs, H., Brugger, E., Whitlock, B., Meredith, J., Ahern, S., Pugmire, D.,
  Biagas, K., Miller, M., Harrison, C., Weber, G.H., Krishnan, H., Fogal, T.,
  Sanderson, A., Garth, C., Bethel, E.W., Camp, D., R\"{u}bel, O., Durant, M.,
  Favre, J.M., Navr\'{a}til, P.: {VisIt: An End-User Tool For Visualizing and
  Analyzing Very Large Data}.
\newblock In: {High Performance Visualization--Enabling Extreme-Scale
  Scientific Insight}, pp. 357--372 (2012)

\bibitem{Dan:2005}
Dan, D., Mueller, C., Chen, K., Glazier, J.A.: Solving the advection-diffusion
  equations in biological contexts using the cellular potts model.
\newblock Phs. Rev. E \textbf{72}, 041,909 (2005)

\bibitem{Davies:1986}
Davies, P.F., Remuzzi, A., Gordon, E.J., Dewey, C.F., Gimbrone, M.A.: Turbulent
  fluid shear-stress induces vascular endothelial-cell turnover in vitro.
\newblock Proc. Natl Acad. Sci. \textbf{83}, 2114--2117 (1986)

\bibitem{DeGroff:2003}
DeGroff, C.G., Thornburg, B.L., Pentecost, J.O., Thornburg, K.L., Gharib, M.,
  \textit{et al.}, D.J.S.: Flow in the early embryonic human heart.
\newblock Pediatric Cardiology \textbf{24}, 375–380 (2003)

\bibitem{Jung:2001}
E.~Jung, C.P.: 2-d simulations of valveless pumping using immersed boundary
  methods.
\newblock SIAM Journal on Scientific Computing \textbf{23}, 19--45 (2001)

\bibitem{Eames:2010}
Eames, S.C., Philipson, L.H., Prince, V.E., Kinkel, M.D.: Blood sugar
  measurement in zebrafish reveals dynamics of glucose homeostasis.
\newblock Zebrafish \textbf{7(2)}, 205–213 (2010)

\bibitem{Forouhar:2006}
Forouhar, A., Liebling, M., Hickerson, A., Nasiraei-Moghaddam, A., Tsai, H.,
  Hove, J., Fraser, S., Dickinson, M., Gharib, M.: The embryonic vertebrate
  heart tube is a dynamic suction pump.
\newblock Science \textbf{312}(5774), 751--753 (2006)

\bibitem{Freund:12}
Freund, J.B., Goetz, J.G., Hill, K.L., Vermot, J.: Fluid flows and forces in
  development: functions, features and biophysical principles.
\newblock Development \textbf{139}, 1229–1245 (2012)

\bibitem{Garita:2011}
Garita, B., Jenkins, M., Han, M., Zhou, C., VanAuker, M., Rollins, A.,
  Watanabe, J., Fujimoto, J., Linask, K.: Blood flow dynamics of one cardiac
  cycle and relationship to mechanotransduction and trabeculation during heart
  looping.
\newblock Am. J. Physiol. Heart Circ. Physiol. \textbf{300}, H879--H891 (2011)

\bibitem{Goenezen:2015}
Goenezen, S., Chivukula, V.K., Midgett, M., Phan, L., Rugonyi, S.: 4d
  subject-specific inverse modeling of the chick embryonic heart outflow tract
  hemodynamics.
\newblock Biomech Model Mechanobiol  (2015)

\bibitem{Granados:2012}
Granados-Riveron, J., Brook, D.: The impact of mechanical forces in heart
  morphogenesis.
\newblock Circ. Cardiovasc. Genet. \textbf{5}, 132--142 (2012)

\bibitem{BGriffithIBAMR}
Griffith, B.: An adaptive and distributed-memory parallel implementation of the
  immersed boundary (ib) method (2014).
\newblock \urlprefix\url{https://github.com/IBAMR/IBAMR}

\bibitem{Griffith:2007}
Griffith, B., Hornung, R., McQueen, D., Peskin, C.: An adaptive, formally
  second order accurate version of the immersed boundary method.
\newblock J. Comput. Phys. \textbf{223}, 10–49 (2007)

\bibitem{Gruber:2004}
Gruber, J., Epstein, J.: Development gone awry-congenital heart disease.
\newblock Circulation Research \textbf{94}, 273--283 (2004)

\bibitem{Hedrick:2008}
Hedrick, T.L.: Software techniques for two- and three-dimensional kinematic
  measurements of biological and biomimetic systems.
\newblock Bioinspiration and Biomimetics \textbf{3}(3), 1--6 (2008)

\bibitem{Hershlag:2011}
Hershlag, G., Miller, L.A.: Reynolds number limits for jet propulsion: a
  numerical study of simplified jellyfish.
\newblock J. Theor. Biol. \textbf{285}, 84--95 (2011)

\bibitem{Heuslein:2014}
Heuslein, J., Meisner, J., Price, R.: Reversal of flow direction enhances
  endothelial cell arteriogenic signaling.
\newblock FASEB J \textbf{28}, 670.9 (2014)

\bibitem{Hove:2003}
Hove, J.R., Koster, R.W., Forouhar, A.S., Acevedo-Bolton, G., Fraser, S.E.,
  Gharib, M.: Intracardiac fluid forces are an essential epigenetic factor for
  embryonic cardiogenesis.
\newblock Nature \textbf{421}(6919), 172--177 (2003)

\bibitem{Howard:2011}
Howard, J., Grill, S.W., Bois, J.S.: Turing's next steps: the mechanochemical
  basis of morphogenesis.
\newblock Nature Reviews Molecular Cell Biology \textbf{12}, 392--398 (2011)

\bibitem{Culver:2010}
J.~Culver, M.D.: The effects of hemodynamic force on embryonic development.
\newblock Microcirculation \textbf{17}, 164--178 (2010)

\bibitem{Lee:2013}
Lee, J., Moghadam, M.E., Kung, E., Cao, H., Beebe, T., Miller, Y., Roman, B.L.,
  Lien, C.L., Chi, N.C., Marsden, A.L., Hsiai, T.K.: Moving domain
  computational fluid dynamics to interface with an embryonic model of cardiac
  morphogenesis.
\newblock PLoS One \textbf{8}, e72,924 (2013)

\bibitem{Liu:2007}
Liu, A., Rugonyi, S., Pentecost, J., Thornburg, K.: Finite element modeling of
  blood flow-induced mechanical forces in the outflow tract of chick embryonic
  hearts.
\newblock Computers and Structures \textbf{85}, 727--738 (2007)

\bibitem{Liu:2010}
Liu, J., Bressan, M., Hassel, D., Huisken, J., Staudt, D., amd K.~Poss, K.K.,
  Mikawa, T., Stainier, Y.: A dual role for erbb2 signaling in cardiac
  trabeculation.
\newblock Development \textbf{137}, 3867--3875 (2010)

\bibitem{Malone:2007}
Malone, M., Sciaky, N., Stalheim, L., Klaus, H., Linney, E., Johnson, G.:
  Laser-scanning velocimetry: A confocal microscopy method for quantitative
  measurement of cardiovascular performance in zebrafish embryos and larvae.
\newblock BMC Biotechnology \textbf{7}, 40 (2007)

\bibitem{Mohammed:2011}
Mohammed, M., Roubaie, S., Jahnsen, E., Jones, E.: Drawing first blood:
  Measuring avian embryonic blood viscosity.
\newblock SURE Poster Presentation \textbf{61}, 33--45 (2011)

\bibitem{Hu:1989}
N.~Hu, E.C.: Hemodynamics of the stage 12 to stage 29 chick embryo.
\newblock Circ. Res. \textbf{65}, 1665–1670 (1989)

\bibitem{Olesen:1988}
Olesen, S.P., Clapham, D.E., Davies, P.F.: Hemodynamic shear-stress activates a
  k+ current in vascular endothelial cells.
\newblock Nature \textbf{331}, 168--170 (1988)

\bibitem{Paluch:2015}
Paluch, E.K., Nelson, C.M., Biais, N., Fabry, B., Moeller, J., Pruitt, B.L.,
  Wollnik, C., Kudryasheva, G., Rehfeldt, F., Federle, W.: Mechanotransduction:
  use the force(s).
\newblock BMC Biology \textbf{13}, 47 (2015)

\bibitem{Patterson:2005}
Patterson, C.: Even flow: Shear cues vascular development.
\newblock Arteriosclerosis, Thrombosis, and Vascular Biology \textbf{25},
  1761–1762 (2005)

\bibitem{Peshkovsky:2011}
Peshkovsky, C., Totong, R., Yelo, D.: Dependence of cardiac trabeculation on
  neuregulin signaling and blood flow in zebrafish.
\newblock Developmental Dynamics \textbf{240}(2), 446?456 (2011)

\bibitem{Peskin:2002}
Peskin, C.: The immersed boundary method.
\newblock Acta Numerica \textbf{11}, 479--517 (2002)

\bibitem{Jamison:2013}
R.~Jamison C.~Samarage, R.B.R.A.F.: In vivo wall shear measurements within the
  developing zebrafish heart.
\newblock PLoS ONE \textbf{8(10)}, e75,722 (2013)

\bibitem{Reckova:2003}
Reckova, M., Rosengarten, C., deAlmeida, A., Stanley, C.P., Wessels, A.,
  Gourdie, R.G., Thompson, R.P., Sedmera, D.: Hemodynamics is a key epigenetic
  factor in development of the cardiac conduction system.
\newblock Circ. Res. \textbf{93}, 77 (2003)

\bibitem{Samsa:2015}
Samsa, L.A., Givens, C., Tzima, E., Didier, Y., Stainer, R., Qian, L., Liu, J.:
  Cardiac contraction activates endocardial notch signaling to modulate chamber
  maturation in zebrafish.
\newblock Development \textbf{142}(6919), 4080--4091 (2015)

\bibitem{Sankova:2010}
Sankova, B., Machalek, J., Sedmera, D.: Effects of mechanical loading on early
  conduction system differentiation in the chick.
\newblock Am. J. Physiol. Heart Circ. Physiol. \textbf{298}, 1571--1576 (2010)

\bibitem{Santhanakrishnan:2011}
Santhanakrishnan, A., Miller, L.: Fluid dynamics of heart development.
\newblock Cell Biochem. Biophys. \textbf{61}, 1--22 (2011)

\bibitem{Santhanakrishnan:2009}
Santhanakrishnan, A., Nguyen, N., Cox, J., Miller, L.: Flow within models of
  the vertebrate embryonic heart.
\newblock J. Theor. Biol. \textbf{259}, 449--461 (2009)

\bibitem{Scherz:2008}
Scherz, P., Huisken, J., Sahai-Hernandez, P., Stainier, D.: High speed imaging
  of developing heart valves reveals interplay of morphogenesis and function.
\newblock Development \textbf{135}, 1179--1187 (2008)

\bibitem{Tarbell:2005}
Tarbell, J.M., Weinbaum, S., Kamm, R.D.: Cellular fluid mechanics and
  mechanotransduction.
\newblock Ann Biomed Eng \textbf{33}, 1719--1723 (2005)

\bibitem{Taylor:1996}
Taylor, A.D., Neelamegham, S., Hellums, J.D., Smith, C.W., Simon, S.I.:
  Molecular dynamics of the transition from l$-$selectin$-$ to beta
  $2-$integrin$-$dependen neutrophil adhesion under defined hydrodynamic shear.
\newblock Biophys. J. \textbf{71(6)}, 3488–3500 (1996)

\bibitem{Tucker:1988}
Tucker, D.C., Snider, C., Jr, W.T.W.: Pacemaker development in embryonic rat
  heart cultured \emph{in oculo}.
\newblock Pediatric Research \textbf{23}, 637--642 (1988)

\bibitem{Turing:1952}
Turing, A.M.: The chemical basis of morphogenesis.
\newblock Proc. R. Soc. Lond. B Biol. Sci. \textbf{237}, 37–72 (1952)

\bibitem{Tytell:2010}
Tytell, E., Hsu, C., Williams, T., Cohen, A., Fauci, L.: Interactions between
  internal forces, body stiffness, and fluid environment in a neuromechanical
  model of lamprey swimming.
\newblock Proc. Natl. Acad. Sci. \textbf{107}, 19,832–19,837 (2010)

\bibitem{Vennemann:2006}
Vennemann, P., Kiger, K.T., Lindken, R., Groenendijk, B.C.W., de~Vos, S.S., ten
  Hagen~\textit{et al.}, T.L.M.: In vivo micro particle image velocimetry
  measurements of blood-plasma in the embryonic avian heart.
\newblock Journal of Biomechanics \textbf{39}, 1191–1200 (2006)

\bibitem{Stekelenburg:2008}
Vos, S.S.D., Ursem, N., Hop, W., Wladimirioff, J., Groot, A.G.D., Poelmann, R.:
  Acutely altered hemodynamics following venous obstruction in the early chick
  embryo.
\newblock J. Exp. Biol. \textbf{206}, 1051--1057 (2003)

\bibitem{Wang:2013}
Wang, C., Baker, B.M., Chen, C.S., Schwartz, M.A.: Endothelial cell sensing of
  flow direction.
\newblock Arterioscler Thromb Vasc Biol. \textbf{33(9)}, 2130--2136 (2013)

\bibitem{Wartlick:2009}
Wartlick, O., Kicheva, A., González-Gaitán, M.: Morphogen gradient formation.
\newblock Cold Spring Harbor Perspectives in Biology \textbf{1}, a001,255
  (2009)

\bibitem{Weinbaum:2003}
Weinbaum, S., Zhang, X., Han, Y., Vink, H., Cowin, S.: Mechanotransduction and
  flow across the endothelial glycoalyx.
\newblock PNAS \textbf{100}, 7988--7995 (2003)

\end{thebibliography}

\end{document}